\documentstyle[seceq,amssymb,graphicx]{jfm}

\ifCUPmtlplainloaded
\else
  \font\tenbmi=cmmib10 at 10pt  \skewchar\tenbmi ='177
  \font\sevenbmi=cmmib10 at 7pt \skewchar\sevenbmi ='177
  \font\fivebmi=cmmib10 at 5pt  \skewchar\fivebmi ='177

  \newfam\bmifam
  \textfont\bmifam=\tenbmi
  \scriptfont\bmifam=\sevenbmi
  \scriptscriptfont\bmifam=\fivebmi
  \def\bmi{\fam\bmifam\tenbmi}
\fi

\ifCUPmtlplainloaded
  \font\sfs = mtssi10 at 10.5pt   
  \font\sls = mtssbi10 at 10.5pt  
  \font\bit = mtmib10 at 10.5pt \skewchar\bit ='177  
\else
  \font\sfs = cmssi10  
  \font\sls = cmssi10  
  \font\bit = cmmib10 \skewchar\bit ='177  
\fi

\ifCUPmtlplainloaded
\else
  \font\eurmten=eurm10
  \font\eurmseven=eurm10 at 7pt
  \font\eurmfive=eurm10 at 5pt
  \newfam\eurmfam
  \textfont\eurmfam=\eurmten
  \scriptfont\eurmfam=\eurmseven
  \scriptscriptfont\eurmfam=\eurmfive
  \edef\eurm{\hexnumber\eurmfam}

  \mathchardef\upi="0\eurm19      
  \mathchardef\umu="0\eurm16      
  \mathchardef\upartial="0\eurm40 

  \font\msxten=msam10
  \font\msxseven=msam10 at 7pt
  \font\msxfive=msam10 at 5pt
  \newfam\msxfam
  \textfont\msxfam=\msxten
  \scriptfont\msxfam=\msxseven
  \scriptscriptfont\msxfam=\msxfive
  \edef\msx{\hexnumber\msxfam}

  \mathchardef\leqslant="3\msx36
  \mathchardef\geqslant="3\msx3E

  \let\leq=\leqslant
  \let\geq=\geqslant

\fi

\newcommand{\be}{\begin{equation}}
\newcommand{\ee}{\end{equation}}
\newcommand{\lb}{\label}
\newcommand{\ol}{\overline}

\newcommand{\bk}{{\bmi k}}
\newcommand{\br}{{\bmi r}}

\newcommand{\bu}{{\bmi u}}
\newcommand{\bv}{{\bmi v}}
\newcommand{\bw}{{\bmi w}}
\newcommand{\bx}{{\bmi x}}

\newcommand{\bD}{\mbox{\sls D}}
\newcommand{\bS}{\mbox{\sls S}}
\newcommand{\bT}{\mbox{\sls T}}

\newcommand{\sS}{\mbox{\sfs S}}
\newcommand{\sT}{\mbox{\sfs T}}

\newcommand{\bdot}{{\mbox{\boldmath $\cdot$}}}
\newcommand{\grad}{{\mbox{\boldmath $\nabla$}}}
\newcommand{\bzed}{{\mbox{\boldmath ${\rm o}$}}}
\newcommand{\bomega}{{\mbox{\boldmath $\omega$}}}
\newcommand{\ben}{{\mbox{\boldmath $\epsilon$}}}

\newcommand{\bvrho}{{\mbox{\boldmath $\varrho$}}}
\newcommand{\btau}{{\mbox{\boldmath $\tau$}}}
\newcommand{\bdots}{{\mbox{\boldmath $:$}}}
\newcommand{\btimes}{{\mbox{\boldmath $\times$}}}

\newcommand{\varPi}{\mathit{\Pi}}
\newcommand{\varGamma}{\mathit{\Gamma}}

\long\def\symbolfootnote[#1]#2{\begingroup%
\def\thefootnote{\fnsymbol{footnote}}\footnote[#1]{#2}\endgroup}

\begin{document}
\title[2D Inverse Energy Cascade]
{A Turbulent Constitutive Law for the Two-Dimensional Inverse Energy Cascade}
\author[Gregory L. Eyink]
{G\ls R\ls E\ls G\ls O\ls R\ls Y\ns L.\ns E\ls Y\ls I\ls N\ls K}
\affiliation{Department of Applied Mathematics \& Statistics,
The Johns Hopkins University, Baltimore, MD 21218}

\date{ }
\maketitle
\begin{abstract}
The inverse energy cascade of two-dimensional (2D) turbulence is
often represented phenomenologically by a Newtonian stress-strain
relation with  a `negative eddy-viscosity'. Here we develop a
fundamental approach to a turbulent constitutive law for the 2D
inverse cascade, based upon a convergent multi-scale gradient
(MSG) expansion. To first order in gradients we find that the
turbulent stress generated by small-scale eddies is proportional
not to strain but instead to `skew-strain,'  i.e. the strain
tensor rotated by $45^\circ.$ The skew-strain from a given scale of
motion makes no contribution to energy flux across eddies at that
scale, so that the inverse cascade cannot be strongly scale-local.
We show that this conclusion extends a result of Kraichnan for
spectral transfer and is due to absence of vortex-stretching in
2D. This `weakly local' mechanism of inverse cascade requires a
relative rotation between the principal directions of strain at
different scales and we argue for this using both the dynamical
equations of motion and also a heuristic model of `thinning' of
small-scale vortices by an imposed large-scale strain. Carrying
out our expansion to second-order in gradients, we find two
additional terms in the stress that can contribute to energy
cascade. The first is a Newtonian stress with an `eddy-viscosity'
due to differential strain-rotation, and the second is a tensile
stress exerted along vorticity contour-lines. The latter was
anticipated by Kraichnan for a very special model situation of
small-scale vortex wave-packets in a uniform strain field. We
prove a proportionality in 2D between the mean rates of
differential strain-rotation and of vorticity-gradient stretching,
analogous to a similar relation of Betchov for 3D. According to
this result the second-order stresses will also contribute to
inverse cascade when, as is plausible, vorticity contour-lines
lengthen on average by turbulent advection.

\end{abstract}

\section{Introduction}

Almost forty years ago, \cite{Kraichnan67} predicted an inverse cascade
of energy in two-dimensional (2D) incompressible fluid turbulence. This is
perhaps
one of the most intriguing turbulent phenomena to occur in classical fluids.
Kraichnan
proposed an inertial range with a $k^{-5/3}$ power-law energy spectrum, just as
in
three dimensions (3D), but with a flux of energy from small-scales to
large-scales rather
than the reverse.
Kraichnan's detailed predictions for steady-state forced 2D turbulence have
been
confirmed  with increasing precision in a series of numerical simulations
[\cite{Lilly71,Lilly72,Fyfeetal77,SiggiaAref81,Hossainetal83,FrischSulem84,
HerringMcWilliams85,MaltrudVallis91,Boffettaetal00}] and laboratory experiments
[\cite{Sommeria86,ParetTabeling98,Rutgers98,Rivera00}]. In fact, it can be
rigorously
proved that an inverse cascade with constant (negative) flux of energy must
occur
in a forced 2D fluid, if damping at low-wavenumbers keeps the energy finite
in the high Reynolds number limit (\cite{Eyink96}). Kraichnan's seminal concept
of an `inverse
cascade' has since been fruitfully extended to other physical situations, such
as inverse
cascade of magnetic helicity in 3D magnetohydrodynamic turbulence
(\cite{Frischetal75}),
of wave action in weak turbulence (\cite{ZakharovZaslavskii82}; see also
\cite{Zakharov67})
and of passive scalars in compressible fluid turbulence
(\cite{Chertkovetal98}).

Attempts have often been made to account for the 2D inverse energy cascade
phenomenon
by a {\it negative eddy-viscosity}, either within analytical closure theories
(\cite{Kraichnan71a,Kraichnan71b,Kraichnan76}) or more phenomenologically
(\cite{Starr68}).
Such a description postulates a constitutive law for the turbulent stress
proportional
to the strain, $\tau_{ij}=-2\nu_T \sS_{ij},$ with a viscosity coefficient
$\nu_T<0$.
However, an exact elimination of turbulent small-scales gives rise to a stress
formula
which is quite different: nonlocal in space, history-dependent and stochastic
(\cite{Lindenberget87,Eyink96b}).  Thus, any local and deterministic
parameterization
of the stress, such as by an eddy-viscosity, can be only an approximate
representation
at best. Nevertheless, such simplified constitutive relations can be quite
useful
to illuminate some of the basic physics of turbulent cascades and they are also
important,
of course, for use in practical numerical modelling schemes.

In a previous paper (Eyink, submitted), hereafter referred to as
(I), we developed a general approximation scheme for the turbulent
stress, based upon a multi-scale gradient (MSG) expansion. We
employed there the {\it filtering approach} to space-scale
resolution in turbulence (\cite{Germano92}), which is also used in
Large-Eddy Simulation (LES) modeling schemes
(\cite{MeneveauKatz00}). Within that framework we developed an
expansion of the stress, first in contributions from different
scales of motion and then in terms of space-gradients of the
filtered velocity field. As a concrete application of the general
scheme we considered in (I) the forward cascade of energy and
helicity in 3D. In this paper we apply the same formalism to the
2D inverse energy cascade. In certain respects, the 2D theory is
more difficult than 3D, because of certain peculiarities of the
inverse cascade. We find that contributions to the stress from
velocity-increments at sub-filter scales are much more important
in 2D that in 3D. Also, terms second-order in space-gradients play
a significant role in the 2D inverse cascade, whereas in 3D the
terms first-order in gradients appear to suffice. Recognizing
these facts has proved crucial to unravelling the physics of the
2D inverse energy cascade.

However, 2D is simpler than 3D in respect of geometry. As we
discussed in (I), the local energy flux is given in general by a
scalar product \be \varPi = -\ol{\bS}\bdots
\btau^{{\,\!}^{\!\!\!\!\circ}} \lb{energy-flux} \ee where
$\ol{\bS}$ is the filtered strain tensor and
$\btau^{{\,\!}^{\!\!\!\!\circ}}$ is the deviatoric (i.e. traceless
part of the) stress tensor $\btau$. The quantity $\varPi$ defined
in (\ref{energy-flux}) represents the rate of work done by the
large-scale strain against the stress induced by the small-scales.
In 3D, this expression involves three eigenvalues for each tensor,
and also three Euler angles which specify the relative
orientations of the tensor eigenframes. However, in 2D one has
simply \be \varPi = -\ol{\sigma}(\delta\tau)\cos(2\theta),
\lb{flux-angle} \ee where $\pm\ol{\sigma}$ are the two eigenvalues
of $\ol{\bS},$ $\pm\delta\tau/2$ are the two eigenvalues of
$\btau^{{\,\!}^{\!\!\!\!\circ}},$ and $\theta$ is the angle
between the eigenframes of these tensors. We have taken $0\leq
\theta\leq \pi/2$ and $\ol{\sigma},\delta\tau\geq 0.$ Thus, the
essence of the inverse energy cascade lies exactly in the tendency
that $0\leq \theta<\pi/4.$  If ${\bmi e}_\pm^{(\tau)}$ are the two
eigenvectors of the deviatoric stress corresponding to the
eigenvalues $\pm\delta\tau/2,$ then there is a net tensile or
expansive stress $\delta\tau/2$ along the ${\bmi e}_+^{(\tau)}$
direction and a net contractile or compressive stress
$-\delta\tau/2$ along the ${\bmi e}_-^{(\tau)}$ direction.
Therefore, when $0\leq \theta<\pi/4$ holds, the stretching
direction ${\bmi e}_+^{(\sigma)}$ of the strain is aligned
primarily along the direction of net tensile stress, whereas the
squeezing direction ${\bmi e}_-^{(\sigma)}$ of the strain is
aligned mainly along the direction of contractile stress. In that
case, the stress cooperates with the strain rather than resists
it, and negative work is done by the large-scales against the
small-scales.

Our primary objective in this work is to gain some understanding how this
characteristic alignment comes about in 2D. In a negative-viscosity model,
the stress is directly proportional to the strain or, equivalently, the
alignment angle $\theta=0.$ This configuration leads to maximal inverse cascade
but it is unlikely to occur uniformly throughout the flow. In fact, a main
result of our work is that, to first-order in gradients and considering
only the contribution to stress from scales of motion near the filter scale,
the alignment is instead $\theta=\pi/4$ everywhere [Section 2.1.1]. We call
such
a stress law `skew-Newtonian' and, from (\ref{flux-angle}), it gives zero
energy flux. Thus, to first-order in gradients, no energy flux can arise in 2D
from strongly scale-local interactions, in agreement with a conclusion
of \cite{Kraichnan71b}. On the other hand, skew-Newtonian stress
from smaller subscale modes can give rise to non-vanishing flux, since the
stress
is oriented at angle $\pi/4$ with respect to the strain at the same scale, not
the large-scale strain $\ol{\bS}$ [Section 2.1.2]. We argue that the flux from
skew-Newtonian stress produced by more distant subfilter scales is negative,
on average, because of a relative rotation of the principal directions
of strain at distinct scales. A plausible explanation for this characteristic
rotation is advanced based on the exact equation for the rotation angle
[Appendix A] and a heuristic model of `vortex-thinning' [Section 2.1.3].
Furthermore, two additional main mechanisms of inverse cascade are predicted
by carrying our expansion to second-order in gradients: a Newtonian stress
with eddy-viscosity due to differential strain-rotation and a tensile stress
directed along vorticity contour-lines [Section 2.2.1]. The latter effect was
anticipated by \cite{Kraichnan76} [Appendix B] and it produces inverse
cascade when vorticity-gradients are stretched by the large-scale strain. We
derive
an identity [Appendix C] that shows that, under the same condition, the
eddy-viscosity due to differential strain-rotation is negative on average
and produces inverse cascade. These mechanisms operate for stress produced
by subfilter scales also, but more weakly the more distant in scale
[Section 2.2.2]. \\

\vspace{-5pt}

\section{The Multi-Scale Model in 2D}

In this section we shall develop for 2D the MSG expansion of the
turbulent stress that was elaborated in general in (I). To keep
our discussion as brief as possible, we shall refer to (I) for
most of the technical details and only outline here the main
points of the general scheme. We employ the standard `filtering
approach' (\cite{Germano92}), which is reviewed, for example, by
\cite{MeneveauKatz00}. Thus, we filter the velocity field $\bu$
with a kernel $G$ at a selected length-scale $\ell$ in order to
define a `large-scale' field $\ol{\bu}$ from scales $>\ell$ and a
complementary `small-scale' field $\bu^\prime=\bu-\ol{\bu}$ from
scales $<\ell.$ However, we further decompose the velocity field
using test kernels
$\varGamma_n(\br)=\ell_n^{-d}\varGamma(\br/\ell_n)$ into
contributions $\bu^{(n)}$ from length-scales
$>\ell_n=\lambda^{-n}\ell.$ The difference $\bu^{[n]}\equiv
\bu^{(n)}-\bu^{(n-1)}$ then represents the velocity contribution
from length-scales between $\ell_{n-1}$ and $\ell_n$ and yields a
multi-scale decomposition
\be \bu = \sum_{n=0}^\infty \bu^{[n]} \ee
of the velocity field. In this paper we assume a scale-ratio
$\lambda=2.$ We also assume, for simplicity, that the kernels $G$
and $\varGamma$ are equal. Thus, the two filtered fields
$\ol{\bu}$ and $\widetilde{\bu}=\bu^{(0)}$ at length $\ell$ are
equal and we need no longer keep the second as a distinct object.

Since the filtered velocity fields $\bu^{(n)}$ are smooth, they
may be Taylor-expanded into a series of terms from $m$th-order
gradients $\grad^m\bu^{(n)}.$ Appropriate functionals of the
velocity field may be expressed in this manner as a summation over
both the gradient index $m$ and the scale index $n$, which we call
a multi-scale gradient (MSG) expansion. Among the most important
quantities for which such a MSG representation may be developed
is the turbulent stress tensor $\btau$. The latter quantity is
defined mathematically as $\btau=\ol{\bu\bu}- \ol{\bu}\,\ol{\bu}.$
Physically, it gives the contribution of the small-scales to
spatial transport of large-scale momentum and it is the quantity
which requires `closure' in the equation for the large-scale
velocity $\ol{\bu}.$ It was proved in (I) that there is a
convergent MSG expansion for the stress tensor, under realistic
conditions for turbulent cascades.

We should remark that two related but distinct approximations for
the subscale stress were developed in (I). The first (I, Section
3) was a systematic expansion, which we shall refer to simply as
the MSG expansion. This is a doubly-infinite series in orders of
space-gradients and in scales of the velocity field, which
converges to the exact subscale stress. However, as discussed in
(I), the rate of convergence of the expansion in order $m$ of
space-gradients is apt to be quite slow as the scale-index $n$ is
increased. To obtain a more rapidly convergent gradient-expansion
in the small-scales, we developed also a more approximate method
(I, Section 4). In this modified approach the small-scale stress
was estimated from velocity-increments for separation vectors in a
certain subset for which the gradient-expansion is rapidly
convergent, at all scales. The hypothesis underlying this
approximation is that the stress due to velocity-increments for
separation vectors from all subregions is similar and can be
estimated, to a good approximation, by the stress arising from the
distinguished subset. We referred to this modified expansion in
(I) as the {\it Coherent-Subregions Approximation} (CSA), or the
CSA-MSG expansion. It is guaranteed to converge rapidly, but its
accuracy depends upon the quality of the basic hypothesis. The
latter seems plausible but should be subjected to empirical tests.

As we shall see below, it is more important to consider the
contributions of subfilter scales in the 2D inverse energy cascade
than it is in the 3D forward cascade. Therefore, the rapid
convergence of the CSA-MSG expansion at small-scales makes it more
practical than the systematic expansion for 2D, and only the
former will be considered here. However, given the close formal
relation between them, most of our qualitative, physical
discussion below can be carried over, with some minor changes, to
the systematic MSG expansion, and it is only for the purpose of
quantitative comparisons that the CSA expansion is to be
preferred. To describe this approximation scheme it is necessary
to decompose the turbulent stress as
$\btau=\bvrho-\bu^\prime\bu^\prime,$ where we refer to $\bvrho$ as
the `systematic' contribution to the stress and to
$-\bu^\prime\bu^\prime$ as the `fluctuation' contribution.
For further discussion of these two terms and for mathematical
formulas, see (I; 2.13-14). In terms of these two quantities,
the general CSA-MSG expression for the stress in any dimension
$d$ was given in (I), to $n$th-order in scale index and
$m$th-order in gradients, as:
\be \btau_*^{(n,m)} = \sum_{k=0}^n \bvrho_*^{[k],(m)}
-\sum_{k,k'=0}^n \bu_*^{\prime\,[k],(m)} \bu_*^{\prime\,[k'],(m)}.
\lb{MNL-stress-II} \ee
Using the results for $m=2$ as illustration, as in (I), we have
\begin{eqnarray}
\bvrho_*^{[k],(2)} & = &
              \frac{\ol{C}_2^{[k]}}{d}\ell^2_k\frac{\partial
\bu^{[k]}}{\partial x_l}
              \frac{\partial \bu^{[k]}}{\partial x_l}
      + \frac{\ol{C}_4^{[k]}}{2d(d+2)}\ell^4_k \frac{\partial^2
\bu^{[k]}}{\partial x_l\partial x_m}
                                    \frac{\partial^2 \bu^{[k]}}{\partial
x_l\partial x_m} \cr
     &  & \,\,\,\,\,\,\,\,\,\,\,\,\,\,\,\,\,\,\,\,
     + \frac{\ol{C}_4^{[k]}}{4d(d+2)}\ell^4_k\bigtriangleup
\bu^{[k]}\bigtriangleup \bu^{[k]}
\lb{stress-diag}
\end{eqnarray}
and
\be \bu_*^{\prime\,[k],(2)} = \frac{-1}{2d\sqrt{N_k}}
             \ol{C}_2^{[k]}\ell^4_k \bigtriangleup \bu^{[k]}
\lb{stress-offdiag} \ee
The coefficients $\ol{C}_p^{[k]}$ in this model for $p=2,4,...$
represent the partial $p$th-moments of the filter-kernel $G$ over
a spherical shell of separation vectors of length $\approx
\ell_k,$ corrected by a multiplicative factor of $N_k=2^{kd}$ to
compensate for the decreasing volume of those shells with
increasing $k.$  Explicit expressions were given for these
coefficients with a Gaussian filter, in (I), Appendix C
\footnote{The expressions involve incomplete Gamma functions.
For the case $d=2$ relevant here, these become, for $p=2m,$
$\gamma\left(\frac{d+p}{2},x\right) = \gamma(1+m,x) = m!
\left[1-\left(1+x+\frac{x^2}{2!}+\cdots
                       +\frac{x^m}{m!}\right)e^{-x}\right],$
in terms of elementary functions. See \cite{AbramSteg}, formulas
6.5.2 and 6.5.13}. Notice that, with the volume-corrected
coefficients used here, the `fluctuation' terms in
(\ref{MNL-stress-II}) are decreased relative to the `coherent'
terms by the factors $1/\sqrt{N_kN_{k'}}.$ These were proposed
in (I) as a consequence of a central limit theorem argument
for the averages over volume that define the `fluctuation'
velocities in (\ref{stress-offdiag}). Because of this, those
terms are expected for larger $k$ to be negligible relative
to the `systematic' contributions in (\ref{MNL-stress-II}).

This brief synopsis provides enough background on the MSG
expansion for our application in this paper to the 2D inverse
cascade. For mathematical derivations and more extensive physical
discussion, see (I).

\subsection{The First-Order Model}

To begin our discussion of the 2D energy cascade, we shall consider
the CSA-MSG expansion of the stress developed to first-order in
velocity-gradients. According to the general formula in equations
(\ref{MNL-stress-II})-(\ref{stress-offdiag}), the expansion of the
stress then contains only the `coherent' part $\bvrho,$ since the
`fluctuation' velocity $\bu^\prime$ vanishes to first-order. Thus,
in any space dimension $d,$ the expansion is given to this order
by
\be \btau_*^{(n,1)} = \sum_{k=0}^n \bvrho^{[k],1}_*
\lb{MNL-stress-I} \ee
with
\be \bvrho_*^{[k],(1)} = \frac{\ol{C}_2^{[k]}}{d}\ell^2_k
\frac{\partial \bu^{[k]}}{\partial x_l}\frac{\partial
\bu^{[k]}}{\partial x_l}, \lb{tau-k-1-anyD} \ee
consisting of just the first term in (\ref{stress-diag}). See also
(I;\,5.2). Terms for large values of $k$ become negligibly small
(UV scale-locality), so that the limit as $n\rightarrow\infty$
exists. For a monofractal velocity field with H\"{o}lder exponent
everywhere $1/3$---as expected in the 2D inverse cascade
(\cite{ParetTabeling98,Yakhot99,Boffettaetal00})---the $k$th term
in (\ref{tau-k-1-anyD}) scales as $\sim \ell_k^{2/3}$
(\cite{Eyink05}and \cite{Eyink05a}).

We now specialize the model to 2D, using the standard formula for
a velocity-gradient (deformation) matrix in 2D,
\be \frac{\partial u_i}{\partial x_j} = \sS_{ij}-\frac{1}{2}
                           \epsilon_{ij}\omega, \lb{gradu-S-om} \ee
which relates it the symmetric, traceless strain matrix $\bS$  and
(pseudo)scalar vorticity $\omega.$ Here $\epsilon_{ij}$ is the
antisymmetric Levi-Civita tensor in 2D. Substituting
(\ref{gradu-S-om}) into (\ref{tau-k-1-anyD}) yields
\be
\varrho_{*;ij}^{[k],1} =
\frac{1}{2}\ol{C}_2^{[k]}\ell^2_k\left\{\sS_{il}^{[k]}\sS_{jl}^{[k]}
+\omega^{[k]}\widetilde{\sS}_{ij}^{[k]}+\frac{1}{4}\delta_{ij}|
\omega^{[k]}|^2\right\} \lb{stress-1stNL2D} \ee
where we have defined the {\it skew-strain matrix} as
$\widetilde{\sS}_{ij}= \sS_{ik}\epsilon_{kj}$\footnote{This
differs slightly from the general definition given in (I), which
would lead us in 2D to term as `skew-strain' instead the product
$\omega^{[k]}\widetilde{\bS}^{[k]}.$ This slight difference in
terminology should cause no difficulty.}. In terms of matrix
arrays \be \bS = \left(\begin{array}{cc}
                    \sS_{11} & \sS_{12} \cr
                    \sS_{12} & -\sS_{11}
                    \end{array}\right), \,\,\,\,\,
\widetilde{\bS} = \left(\begin{array}{cc}
                             -\sS_{12} & \sS_{11} \cr
                              \sS_{11} & \sS_{12}
                             \end{array}\right).  \lb{S-skewS} \ee
Thus, the skew-strain is also symmetric and traceless. It is easy
to see that the strain and skew-strain are orthogonal in the
standard matrix inner-product $\bS\bdots\,\widetilde{\bS}=0$ (and
hence the prefix `skew'). The various terms that appear in
(\ref{stress-1stNL2D}) are the same as those in equation (I;\,5.4)
for 3D and have the same physical interpretations. Note, however,
a principal difference with 3D is the absence of terms
proportional to $\omega_i^{[k]}\omega_j^{[k]}.$ Since the only
component of vorticity is perpendicular to the plane of motion,
no stress can be directed along vortex-lines in 2D.

\subsubsection{The Strong UV-Local Terms}

It is interesting to consider separately the first term in
(\ref{MNL-stress-I}), for $k=0,$ since it corresponds to the
stress contribution from filter-scale velocity-increments. Thus,
we refer to this as the strongly UV-local contribution. It is the
only summand in the formula (\ref{MNL-stress-I}) which is closed
in terms of the filtered velocity $\ol{\bu}=\bu^{(0)}.$ In fact,
this term corresponds just to the well-known Nonlinear Model for
the turbulent stress (\cite{MeneveauKatz00}), as discussed at
length in (I).

The most important observation about the strongly UV-local term in
2D is that it gives zero energy flux, pointwise in space. This is
obvious for the term proportional to $|\ol{\omega}|^2,$ since it
is a pressure contribution. Furthermore, the first term is
proportional to $\ol{\bS}^2=\ol{\sigma}^2{\bmi I}$ in 2D, where
${\bmi I}$ is the identity matrix, and is thus also a pressure
contribution. Here we have used the Cayley-Hamilton theorem and
the fact that the strain matrix in 2D has two eigenvalues $\pm
\ol{\sigma}$ of equal magnitude but opposite sign. Therefore, the
first term contributes also zero flux. The term in
(\ref{stress-1stNL2D}) proportional to the skew-strain is
deviatoric but it does not contribute to energy flux, by the
orthogonality mentioned earlier. We can thus conclude that there
is no energy flux anywhere in space arising from the strongly
UV-local interactions, to first-order in velocity-gradients.

This conclusion agrees with a result of \cite{Kraichnan71b}, who
showed that energy cascade in 2D cannot be strongly scale-local.
It is worthwhile to summarize his demonstration, which is based on
the detailed conservation of energy and enstrophy in Fourier
space. Let $T(k,p,q)$ represent the energy transfer into
wavenumber magnitude $k$ from all triads of wavenumbers with
magnitudes $k,p,q.$ A measure of the scale-locality of the triad
is provided by the parameter
$$\nu=\log_2(k_{{\rm med}}/k_{\min})\geq 0, $$
where $k_{\min},k_{{\rm med}},k_{\max}$ are the minimum, median,
and maximum wavenumber magnitudes, respectively, from the triad
$k,p,q$. Intuitively, this quantity represents the `number of
cascade steps' between the minimum and median wavenumber. Note
that $k_{\max}\leq 2k_{{\rm med}}$ by wavenumber addition, so that
$\log_2(k_{\max}/k_{\min})\leq \nu+1.$ Thus, the parameter $\nu$
unambigously measures the ratio of scales involved in the triadic
interaction. In these terms, nonlocal\footnote{Note that this
definition makes no distinction between ultraviolet (UV) and
infrared (IR) nonlocal interactions, as in \cite{Eyink05}.}
interactions correspond to those triads with $\nu\gg 1$ and
strongly scale-local ones to those with $\nu\ll 1.$
\cite{Kraichnan71b} noted that in 2D the transfer function
satisfies both
\be T(k,p,q)+T(p,q,k)+T(q,k,p)=0 \lb{detailed-energy} \ee
as a consequence of energy conservation, and
\be k^2T(k,p,q)+p^2T(p,q,k)+q^2T(q,k,p)=0
\lb{detailed-enstrophy} \ee
by conservation of enstrophy. Multiplying through
(\ref{detailed-energy}) by $q^2$ and subtracting from
(\ref{detailed-enstrophy}) gives
\be (k^2-q^2)T(k,p,q)+(p^2-q^2)T(p,q,k)=0. \lb{Kraichnan-deriv}
\ee
Thus, if $k\neq p=q,$ then $T(k,p,p)=0,$ and substituting back
into (\ref{detailed-energy}) gives also $T(p,p,k)=T(p,k,p)=0.$
Hence, there is zero transfer, if any two wavenumbers have equal
magnitudes, and, in particular, if $\nu=0.$ However, it is very
plausible to expect that the transfer function will be continuous
in the wavenumber magnitude. In that case, transfer will be
vanishingly small also in the limit that $\nu\ll 1.$
\cite{Kraichnan71b} obtained more quantitative results using his
analytical Test-Field-Model (TFM) closure. He found (see his
Figure 2) that roughly 90\% of the energy flux comes from triads
with $\nu\geq 1,$ 70\% with $\nu\geq 2$, and 60\% with $\nu\geq
3$. To obtain 90\% of the total energy flux in the TFM closure
required including all triads with $\nu\leq 5.$ Thus, the 2D
energy cascade was predicted by the Kraichnan to be scale-local
(cf. also the exact analysis in \cite{Eyink05}) but only weakly
so.

There is a fundamental relationship between our argument and
Kraichnan's. This is best understood by recalling the form of the
energy flux in 3D from the strongly local, first-order terms,
equation (I;\,5.11) [and see also \cite{BorueOrszag98}]:
\be \varPi^{(0,1)} = \frac{1}{3}C_2\ell^2 \left\{-{\rm
Tr}\,(\ol{\bS}^3)
         +\frac{1}{4}\ol{\bomega}^\top\ol{\bS}\ol{\bomega}\right\} \lb{I-5.8}
\ee
Both of these terms vanish in 2D, the second because of absence of
vortex-stretching. As discussed in (I), the first term can also be
related to vortex-stretching, at least in a space-average sense,
by a relation of \cite{Betchov56}. Of course, the lack of
vortex-stretching in 2D is also what underlies the conservation of
enstrophy, used in Kraichnan's argument. The argument that we have
given confirms Kraichnan's conclusion and extends it to be also
pointwise in space.

\subsubsection{The Weakly UV-Local Terms}

{}From the preceding discussion we can see that any energy flux that
arises to first order in gradients must be due to subfilter modes,
with $k>0.$ Since the contribution from modes with $k\gg 1$ is
also small, the flux comes primarily from the weakly local terms
with $k\gtrsim 1.$  This contribution for each $k\geq 1$ can arise
solely from the skew-strain term in the stress
(\ref{stress-1stNL2D}), since, by the same reasoning as above, the
other two terms are isotropic stresses or pressures. The flux from
modes at scale $k$ is thus
\be \varPi^{[k],(1)}_* = \frac{1}{2}\ol{C}_2^{[k]}\ell^2_k
\omega^{[k]}\left(\bS^{(0)}\bdots\,\widetilde{\bS}^{[k]}\right)
\lb{MNL-flux-1st} \ee
This can be rewritten in more intuitive fashion using `polar
coordinates' for strain matrices:
\be \bS = \sigma \left(\begin{array}{cc}
                           \cos(2\alpha) & \sin(2\alpha) \cr
                           \sin(2\alpha) & -\cos(2\alpha)
                    \end{array}\right), \,\,\,\,\, \widetilde{\bS}
         = \sigma \left(\begin{array}{cc}
                             -\sin(2\alpha) & \cos(2\alpha) \cr
                             \cos(2\alpha)  & \sin(2\alpha)
                             \end{array}\right).  \lb{stress-polar} \ee
Here $\sigma=|\bS|/\sqrt{2}$ is the (positive) strain eigenvalue
and $\alpha=(1/2)\arctan(\sS_{12}/\sS_{11})$ is the angle made by
the frame of strain eigenvectors ${\bmi e}_+^{(\sigma)},{\bmi
e}_-^{(\sigma)}$ with a fixed orthogonal frame. Note,
incidentally, that the skew-strain is obtained by rotating the
frame of the strain by $\pi/4$ radians. By choosing appropriately
among the two unit eigenvectors $\pm{\bmi e}_+^{(\sigma)},$ one
can always ensure that $0\leq |\alpha|<\pi/2.$  Thus, from
(\ref{MNL-flux-1st}), (\ref{stress-polar}),
\be \varPi^{[k],(1)}_* = \ol{C}_2^{[k]}\ell^2_k
\sigma^{(0)}\sigma^{[k]}
\omega^{[k]}\sin[2(\alpha^{[k]}-\alpha^{(0)})],
\lb{MNL-flux-1st-polar} \ee
a remarkably simple and compact result.

The total flux from all scales $k=0,1,...,n$ to first order in
gradients is thus
\be \varPi^{(n,1)}_*
        = \sum_{k=1}^n \ol{C}_2^{[k]}\ell^2_k \sigma^{(0)}\sigma^{[k]}
                       \omega^{[k]}\sin[2(\alpha^{[k]}-\alpha^{(0)})].
\lb{MNL-total-flux-1st-polar} \ee
In order to achieve an inverse energy cascade, it must hold that
the terms in the sum are negative on average, at least for
$k\gtrsim 1.$ The sign of (\ref{MNL-flux-1st-polar}) is determined
completely by the factor
$\omega^{[k]}\sin[2(\alpha^{[k]}-\alpha^{(0)})],$ which depends
upon the relative angle $\alpha^{[k]}-\alpha^{(0)}.$ If we choose
that $0\leq |\alpha^{[k]}-\alpha^{(0)}|<\pi/2,$ then this factor
will be negative if the strain-frame at scale $k$ lags the
strain-frame at scale $0$ ($\alpha^{[k]}<\alpha^{(0)}$) in regions
where $\omega^{[k]}>0,$ and leads ($\alpha^{[k]}>\alpha^{(0)}$) in
regions where $\omega^{[k]}<0.$ Under these conditions, the
small-scale stress will cooperate with the large-scale strain and
the latter will do negative work. Note that this is quite
different from a `negative-viscosity' mechanism, with a Newtonian
stress proportional to strain $\bS^{[k]}.$ Instead, the crucial
deviatoric component of the stress is of the form
$\gamma^{[k]}\widetilde{\bS}^{[k]},$ where $\gamma^{[k]}=
\ol{C}_2^{[k]}\ell^2_k \omega^{[k]}/2$ has dimensions of a diffusion
constant and may be termed `skew-viscosity.'

We see that a contribution to inverse energy cascade at scale $k$
requires an anti-correlation in the signs of $\omega^{[k]}$ and
$\alpha^{[k]}-\alpha^{(0)}.$ A plausible dynamical mechanism for
this can be suggested, based upon the exact equation for the
strain orientation angle:
\be (\sigma^{(k)})^2
D_t^{(k)}\alpha^{(k)} = \frac{1}{8}\omega^{(k)}Q^{(k)}
            + \grad\bdot {\bmi K}^{(k)} + \cdots,
\lb{alpha-n-dot} \ee
with $D_t^{(k)}=\partial_t + \bu^{(k)}\bdot\grad$ the advective
derivative at scale $k,$
\be  Q^{(k)}=\bigtriangleup p^{(k)} = \frac{1}{2}[\omega^{(k)}]^2
-2[\sigma^{(k)}]^2  \lb{press-hess-n} \ee
the pressure hessian at scale $k,$ and
\be  {\bmi K}^{(k)} = \frac{1}{4} (\grad p^{(k)}\btimes
\grad)\bu^{(k)} \lb{K-curr-n} \ee
a space transport term due to pressure forces. The $\cdots$ terms
in (\ref{alpha-n-dot}) represent contributions from the turbulent
stress due to modes at length-scales $<\ell_k.$ See Appendix A for
the derivation. According to (\ref{alpha-n-dot}),
$D_t{\alpha}^{(k)} \sim \ell_k^{-2/3}$ in a 2D inverse energy
cascade range, so that the rotation-rate increases with increasing
$k$. Since the pressure contribution $\grad\bdot {\bmi K}^{(k)}$
is spatially-nonlocal and averages to zero, it can be treated as
random noise. We shall likewise disregard the effect of subgrid
terms $\cdots.$  Thus, the expected correlation will be created in
strain-dominated regions with $Q^{(k)}<0,$ since $\alpha^{(k)}$
there rotates against the locality vorticity $\omega^{(k)}$ and
faster for larger $k.$ Since the flux (\ref{MNL-flux-1st-polar})
is proportional also to the strain magnitudes
$\sigma^{(0)},\sigma^{(k)},$ most of the cascade should occur in
the strain regions where this counter-rotation occurs.

\subsubsection{A Heuristic Model}

A simple model problem may help to illuminate the basic mechanism
of inverse energy cascade due to skew-strain. We shall consider
the effect of a large-scale uniform straining field
\be \bS^{(0)}
= \left(
              \begin{array}{cc}
                \sigma^{(0)} & 0 \cr
                0 & -\sigma^{(0)}
              \end{array} \right), \lb{thin-strain} \ee
on a collection of small-scale vortices, each initially circular
with support radius $\ell_n.$ The $i$th vortex in the assembly
will be assumed to have initially a vorticity distribution
$\omega_i^{[n]}(|\br-\br_i|)$ radially symmetric about its center
$\br_i.$ Let us assume also that the small-scale vortices have
each a single sign of vorticity, but with the net circulation of
the array equal to zero: $\sum_i\int d\br
\,\,\omega_i^{[n]}(r)=0.$ \cite{Kraichnan76} considered a very
similar model problem of ``vortex-blobs'' in order to illustrate
the mechanism of asymptotic negative viscosities in his Test-Field
Model closure. In Appendix B we review Kraichnan's ``blob-model''
and compare it with the present one. Suffice it to say here that
it was crucial in Kraichnan's calculation to take vortex
wave-packets with a very rapid sinusoidal variation in the
vorticity. On the contrary, we require no such variation and a
particular case of our model is an array of vortex patches with
constant vorticity levels, each initially circular.

The effect of the straining field on this set of small-scale
vortices will be to deform them into elliptical form, elongated in
the $x$-direction and thinned in the $y$-direction. \cite{Kida81}
found this behavior in his exact solution of 2D Euler for an
elliptical vortex patch in a uniform shear flow, whenever the
strain $\sigma$ and vorticity level $\omega$ satisfy
$|\sigma/\omega|\geq (3-\sqrt{5})/[2(2+2\sqrt{5})^{1/2}]\doteq
0.15.$ More generally, the same phenomenon appears in a rapid
distortion limit for the case of a strong strain $\sigma^{(0)}\gg
\max_i\|\omega^{[n]}_i\|_\infty.$ We can then ignore the
self-evolution of the vortices and also their mutual interactions.
This permits us to focus on a single vortex centered at
$\br=\bzed$ with radial vorticity profile $\omega^{[n]}(r).$  The
vorticity level set initially at radius $r$ is distorted into an
ellipse whose equation is $x^2/a^2+y^2/b^2=1$ with semimajor axis
$a=r\exp[\sigma^{(0)}t]$ and semiminor axis
$b=r\exp[-\sigma^{(0)}t]$ at time $t.$

The immediate result is that the energy of the small-scale vortex
patch is reduced, as a consequence of conservation of circulation.
The area inside each elliptical vorticity contour is preserved,
but the length of the perimeter is increased. In order to keep the
circulation constant, the circumferential velocity must decrease.
For example, in the case of a circular vortex patch of constant
vorticity-level $\omega^{[n]}$ with initial radius $r=\ell_n,$ the
patch evolves into an elliptical shape with circulation
$\omega^{[n]}\cdot \pi ab \doteq 4u^{[n]}a,$ where $u^{[n]}(t)$ is
the $x$-component of the circumferential velocity at time $t$. The
second expression for circulation holds in the limit when
$\sigma^{(0)}t\gg 1$ and $a\gg b,$ so that the perimeter of the
elliptical vortex is approximately $4a$ and is nearly parallel to
the $x$-axis. In that case,
\be u^{[n]}(t) \doteq (\pi/4) \omega^{[n]}b
        =(\pi/4) \omega^{[n]}\ell_n\exp[-\sigma^{(0)}t].
        \lb{thin-circum-vel} \ee
A similar argument can be made for points interior and exterior to
the vortex, with the result that the velocity is everywhere
reduced by a common factor of $\exp[-\sigma^{(0)}t].$ Thus, the
kinetic energy of the vortex is also decreased. (Of course, a
single vortex of definite sign would have infinite energy in the
unbounded plane, due to divergence at infinity. Such far-field
divergence is absent when considering the array of vortices with
zero net circulation.)

\begin{figure}
\centering
\includegraphics[width=150pt,height=200pt]{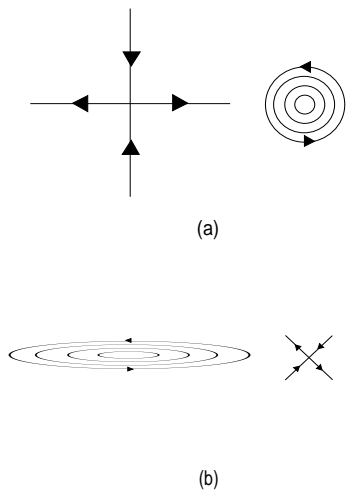}
\caption{{\small Mechanism of vortex-thinning. (a) A large-scale
strain field with stretching direction along the $x$-axis and shrinking
direction along the $y$-axis, and a small-scale vortex of positive
circulation, initially circular; (b) The vortex elongated along the $x$-axis
and thinned along the $y$-axis, and its strain basis, rotated by $-45^\circ$
with respect to the large-scale strain.}}
\label{thinning}
\end{figure}

The energy lost by the collection of small-scale vortices is
transferred to the large-scales. To see this, observe that the
large-scale straining, in addition to reducing the velocity
amplitude of the small-scale vortices, also rectifies the velocity
direction. The velocity vector of  the elongated vortices points
almost entirely in the $x$-direction and very little in the
$y$-direction. Indeed, the vorticity level curve initially at
radius $r$ for the profile $\omega^{[n]}(r)$ now becomes, to
leading order, a pair of straight, parallel lines $y=\pm b=\pm
r\exp[-\sigma^{(0)}t].$ Thus, the vorticity field approximates to
$\omega^{[n]}(y,t)= \omega^{[n]}(|y|\exp[\sigma^{(0)}t])$ when
$\sigma^{(0)}t\gg 1.$ This is just the vorticity associated to a
long, narrow shear layer with weakened velocity
\be u^{[n]}(y,t) = -\exp[-\sigma^{(0)}t]{\rm sign}\,(y)
\int_0^{|y|\exp[\sigma^{(0)}t]} \omega^{[n]}(r)\,dr \lb{thin-veloc}
\ee
directed entirely along the $x$-axis. If the tensor product
$\bu^{[n]}\bu^{[n]}$ were integrated over space at the initial
time, it would produce only a diagonal stress contribution:
\be \sT_{ij}(t=0) =\int_{{\rm vortex}}d\br \,\,u^{[n]}_i(\br)
u^{[n]}_j(\br) = \delta_{ij} \cdot \pi \int_0^{\ell_n} dr
\,\,r|u_\theta^{[n]}(r)|^2, \lb{thin-stress-before} \ee
where
$u_\theta^{[n]}(r)=(1/r)\int_0^r\,\rho\omega^{[n]}(\rho)\,d\rho$
is the tangential velocity around the vortex center. (Here we have
integrated only over the body of the vortex, neglecting the
contribution of more distant regions). However, after
``rectification'' there is a net stress component
\begin{eqnarray}
\sT_{11}(t) & \doteq & 2a\int_{-b}^{b} dy\,
            u^{[n]}(y,t) u^{[n]}(y,t) \cr
          & = & 4\ell_n \int_0^{\ell_n} dr\,
            \left\{\int_0^r \omega^{[n]}(\rho)\,d\rho\right\}^2,
\lb{thin-stress-after} \end{eqnarray}
with all other components much smaller. This resultant stress
reinforces the large-scale strain field, so that $\int
d\br\,\varPi(\br,t)= -\sS_{ij}\sT_{ij}<0,$ and  negative work is
done by the large-scales against the small scales.

This simple model of inverse energy cascade illustrates the
pattern of relative orientation of strain frames at distinct
scales, which was discussed earlier. In fact, within the long,
narrow shear layer created by thinning of a vortex there is a
velocity-gradient (or deformation) tensor of the form
\be \bD^{[n]}(y,t) = \left(
              \begin{array}{cc}
                0 & -\omega^{[n]}(y,t) \cr
                0 &  0
              \end{array} \right), \lb{thin-D} \ee
with $(\partial u^{[n]}/\partial y)(y,t)=-\omega^{[n]}(y,t).$ The
corresponding strain matrix is
\be \bS^{[n]}(y,t) = \left(
              \begin{array}{cc}
                0 & -\omega^{[n]}(y,t)/2 \cr
                -\omega^{[n]}(y,t)/2 &  0
              \end{array} \right), \lb{thin-S} \ee
which has eigenvectors
\be  {\bmi e}_+^{[n]} = \left(
              \begin{array}{c}
                1 \cr
                -1
              \end{array} \right), \,\,\,\,
  {\bmi e}_-^{[n]} = \left(
              \begin{array}{c}
                1 \cr
                1
              \end{array} \right) \lb{thin-eigvecs} \ee
for $\omega^{[n]}(y,t)>0$ and with ${\bmi e}_+^{[n]},{\bmi
e}_-^{[n]}$ reversed for $\omega^{[n]}(y,t)<0.$ See Figure 1,
which illustrates the case of a vortex patch of positive
(counterclockwise) circulation. The small-scale strain basis shown
there is rotated relative to the large-scale strain basis by
$-\pi/4$ radians. If the vortex patch had had negative (clockwise)
circulation, then the rotation would have been by $+\pi/4$ radians
instead.

This same model also clarifies the origin of stress proportional
to skew-strain in our general scheme. The skew-strain in such an
elongated vortex is
\be \widetilde{\bS}^{[n]}(y,t) = \left(
              \begin{array}{cc}
                \omega^{[n]}(y,t)/2 & 0 \cr
                0 & -\omega^{[n]}(y,t)/2
              \end{array} \right). \lb{thin-skew-S} \ee
Let us introduce a convenient space-average over the vortex of the
form
\be  \langle\omega^{[n]}\rangle=(2/b^2)\int_0^b dy\,\,\int_0^y
dy'\,\,\omega^{[n]}(y',t)=(2/\ell_n^2)\int_0^{\ell_n}
dr\,\,\int_0^r d\rho\,\,\omega^{[n]}(\rho). \lb{thin-omega} \ee
By the Cauchy-Schwartz inequality, $ (1/\ell_n) \int_0^{\ell_n}
dr\,
            \left\{\int_0^r \omega^{[n]}(\rho)\,d\rho\right\}^2
            \geq \left[(\ell_n/2)\langle\omega^{[n]}\rangle\right]^2,
$ and, furthermore, these two quantities will generally have a
ratio within some specified bounds. It follows that, when
$\sigma^{(0)}t\gg 1,$
\be \ell_n^2
\langle\omega^{[n]}\rangle\langle\widetilde{\bS}^{[n]}\rangle
   \doteq  \left(
         \begin{array}{cc}
         \tau_{11}/2 & 0 \cr
          0 & -\tau_{11}/2
         \end{array} \right), \lb{thin-skew-tau} \ee
where we have set $\tau_{11}=T_{11}/\ell_n^2.$ Thus, a
(deviatoric) stress proportional to skew-strain arises naturally
from a narrow shear layer produced by vortex-thinning.

It is not completely obvious why small-scale vortices in a
two-dimensional inverse cascade range should be elongated and
thinned by large-scale strain. After all, in such a range
$\sigma^{(0)}\sim \ell^{-2/3}\ll (\ell_n)^{-2/3} \sim
\omega^{[n]}$ for $\ell\gg \ell_n.$ Thus, the large-scale strain
is weak compared with the vorticity at smaller scales, exactly the
opposite as is assumed in the rapid distortion limit above. The
vorticity at length-scale $\ell_n$ could be expected to respond
more strongly to the larger strains $\sigma^{[n']}\gg
\omega^{[n]}$ from length-scales $l_{n'}\ll \ell_n.$ However, the
large-scale strain, although relatively weak, is coordinated over
large distances and is temporally coherent, with a typical
lifetime of $t_\ell \sim \ell^{2/3}.$ By contrast, the strain from
the smaller scales is random and uncoordinated and, furthermore,
evolves on a much shorter time-scale $t_{\ell_{n'}}\sim
(\ell_{n'})^{2/3}.$ Thus, the small-scale vorticity can adjust
very rapidly to the persistent large-scale strain, whereas it does
not have time to adjust to the many, even more rapidly fluctuating
strains from the still smaller scales.

Clearly, our simple model calculation does not reflect all of
the complexities of the two-dimensional inverse cascade range.
However, it gives a simple physical picture for the origin of
stress proportional to skew-strain, which, we believe, is
essentially the correct one. If the initial profiles of the
vorticity, $\omega_i^{[n]}(r)$ for the $i$th circular vortex, are
not constant in the radial distance $r$ from the center, then
vortex-thinning produces also large vorticity-gradients parallel
to the compressing direction of the strain field. This
second-order effect will be discussed in detail in the following
section.


\subsection{The Second-Order Model}

We have seen that, unlike in 3D, the MSG expansion $\btau_*^{(n,m)}$
to lowest order in space-gradients, $m=1,$ can only explain energy cascade
if subfilter scales $n\geq 1$ are considered. However, another possible
mechanism may be terms of higher order in space-gradients with $m\geq 2$.
To investigate this possibility, we develop in this section the 2D MSG
expansion to second-order in velocity-gradients. One can specialize the
formulas (\ref{stress-diag}),(\ref{stress-offdiag}) to 2D, replacing the
velocity derivative with strain and vorticity using (\ref{gradu-S-om}).
The result is:
\begin{eqnarray}
{\varrho}_{*;ij}^{[k],(2)} & = &
\frac{1}{2}\ol{C}_2^{[k]}\ell^2_k\left[\sS_{il}^{[k]}\sS_{jl}^{[k]}
+\omega^{[k]}\widetilde{\sS}_{ij}^{[k]}+\frac{1}{4}\delta_{ij}
|\omega^{[k]}|^2\right] \cr
                      &   & \,\,\,\,\,\,
+\frac{1}{16}\ol{C}_4^{[k]}\ell^4_k\left[\sS_{il,m}^{[k]}\sS_{jl,m}^{[k]}
+(\partial_l\omega^{[k]})\widetilde{\sS}_{ij,l}^{[k]}+\frac{1}{4}
\delta_{ij}|\grad\omega^{[k]}|^2\right] \cr
    \, &   &
\,\,\,\,\,\,\,\,\,\,\,\,\,\,\,\,\,\,\,\,\,\,\,\,
\,\,\,\,\,\,\,\,\,\,\,\,\,\,\,\,\,\,\,\,
+\frac{1}{32}\ol{C}_4^{[k]}\ell^4_k\widetilde{\partial}_i
\omega^{[k]}\,\widetilde{\partial}_j\omega^{[k]}.
\lb{MNL-stress-2D-diag}
\end{eqnarray}
and
\be u_{*;i}^{\prime\,[k],(2)} = \frac{1}{4\sqrt{N_k}} \ol{C}_2^{[k]}\ell^2_k
 \widetilde{\partial}_i\omega^{[k]}. \lb{MNL-stress-2D-off} \ee
In the last term of (\ref{MNL-stress-2D-diag}) and also in
(\ref{MNL-stress-2D-off}) we
have defined $\widetilde{\partial}_i=\epsilon_{ij}\partial_j,$ the {\it
skew-gradient},
which satisfies $\widetilde{\grad}\bdot\grad=0.$ This is the same operator that
appears
in the stream-function representation of a velocity
$u_i=\widetilde{\partial}_i\psi$.
Indeed, to derive the last term in (\ref{MNL-stress-2D-diag}) and the term in
(\ref{MNL-stress-2D-off}) we used the stream function $\psi^{[k]}$ and the
Poisson equation
$-\bigtriangleup\psi^{[k]}=\omega^{[k]}$ in order to write $\bigtriangleup
u_i^{[k]}
=-\widetilde{\partial}_i\omega^{[k]}.$

\subsubsection{The Strongly UV-Local Terms}

As for the first-order expansion, we begin by considering just the strongly
UV-local terms
with $n=0$. These give altogether (note that $C_p^{(0)}=\ol{C}_p^{(0)}$ for
$n=0$)
\begin{eqnarray}
{\varrho}_{*;ij}^{(0,2)} & = &
\frac{1}{2}C_2^{(0)}\ell^2\left[\ol{\sS}_{il}\ol{\sS}_{jl}
+\ol{\omega}\widetilde{\ol{\sS}}_{ij}+\frac{1}{4}\delta_{ij}
\ol{\omega}^2\right] \cr
&   & \,\,\,\,\,\,
+\frac{1}{16}C_4^{(0)}\ell^4\left[\ol{\sS}_{il,m}\ol{\sS}_{jl,m}
+(\partial_k\ol{\omega})\widetilde{\ol{\sS}}_{ij,k}+\frac{1}{4}
\delta_{ij}|\grad\ol{\omega}|^2\right] \cr
    \, &   &
\,\,\,\,\,\,\,\,\,\,\,\,\,\,\,\,\,\,\,\,\,\,\,
\,\,\,\,\,\,\,\,\,\,\,\,\,\,\,\,\,\,\,\,\,
+\frac{1}{32}[C_4^{(0)}-2(C_2^{(0)})^2]\ell^4\,\,\,\,
(\widetilde{\partial}_i\ol{\omega})\,(\widetilde{\partial}_j\ol{\omega}).
\lb{NL-model-2D}
\end{eqnarray}
Let us consider the physical meaning of the various terms that appear.

We have already considered the terms in the initial line of (\ref{NL-model-2D})
that
arise from first-order velocity-gradients and have shown that they give no
contribution
to energy flux. The second line is remarkably similar in appearance to the
first.
In fact, it is not hard to see that the first term proportional to
$\ol{\sS}_{il,m}\ol{\sS}_{jl,m}$ is an isotropic (pressure) term, by exactly
mimicking
the argument we gave earlier for the $\ol{\sS}_{il}\ol{\sS}_{jl}$-term,
separately
for each value of the index $m$ that is summed over. Of course, the final term
proportional to $\delta_{ij}|\grad\ol{\omega}|^2$ is also a pressure. This
leaves
only the middle deviatoric term as possibly contributing to energy flux. It is
interesting that this second-order term,
\be \btau^{(0),[2]}_{I}
= (1/16)C_4^{(0)}\ell^4(\grad\ol{\omega}\bdot\grad)\widetilde{\ol{\bS}}, \ee
gives rise exactly to an `eddy-viscosity'. To see this, it is easiest to use
the
`polar coordinates' (\ref{stress-polar}) for the strain and skew-strain.
Together
with the chain rule, this gives
\be (\grad\ol{\omega}\bdot\grad)\widetilde{\ol{\bS}}=-2(\grad\ol{\omega}\bdot
\grad\ol{\alpha})\ol{\bS}+(\grad\ol{\omega}\bdot
\grad\ol{\lambda})\widetilde{\ol{\bS}}
\lb{Diff-TildeS} \ee
with $\ol{\lambda}=\ln \ol{\sigma}.$ Of course, the second term proportional to
skew-strain does not contribute to energy flux. Thus, up to such conservative
terms,
we obtain
\be \btau^{(0),[2]}_{I} =
-(1/8)C_4^{(0)}\ell^4(\grad\ol{\omega}\bdot\grad\ol{\alpha})\ol{\bS}
    + \cdots = -2\nu_T \ol{\bS}, \ee
with $\nu_T=C_4^{(0)} \ell^4 (\grad\ol{\omega}\bdot\grad\ol{\alpha})/16.$
This is a stress of Newtonian form, with an eddy-viscosity due to
differential-rotation of the strain. Indeed, the eddy-viscosity coefficient
$\nu_T$ is just proportional to the rate of rotation of strain along the
direction of maximum increase of vorticity.

The final term of (\ref{NL-model-2D}) arises from the combination of the last
term in (\ref{MNL-stress-2D-diag}) for $k=0$ and the product of two terms in
(\ref{MNL-stress-2D-off}) for $k=k'=0.$ These together give a stress exerted
along the direction parallel to the skew-gradient
$\widetilde{\grad}\ol{\omega}$.
Equivalently, this stress is directed normal to the vorticity-gradient
$\grad\ol{\omega},$ or along the level-sets or contour-lines of the vorticity.
There are two opposing contributions, a tensile stress proportional to
$C_4^{(0)}$ from (\ref{MNL-stress-2D-diag}) and a contractile stress
proportional to $(C_2^{(0)})^2$ from (\ref{MNL-stress-2D-off}). Which
dominates could depend upon the choice of the filter kernel $G$. However,
the concrete calculations in (I), Appendix C show that
$C=[C_4^{(0)}-2(C_2^{(0)})]^2/32>0$ for a Gaussian kernel. We have also
checked this to be true for a few other cases, e.g. an exponential filter
$G(\br)=e^{-|\br|}/(2\pi).$ At least for these choices we see that there
is a tensile stress of strength $C\ell^4|\grad\ol{\omega}|^2$ exerted
by the small-scales along vorticity contour-lines. As we discuss in
Appendix B of the present paper, this effect was anticipated in a calculation
of \cite{Kraichnan76} for a simple model problem of a 2D vorticity wave-packet
in a uniform strain field. This tensile stress along vorticity contours should
be contrasted with the contractile stress $-C_2^{(0)}\ell^2|\ol{\bomega}|^2/2$
exerted along vortex-lines in 3D, discussed in (I).

The strongly UV-local terms in the stress thus can give a non-vanishing
contribution to energy flux, at second-order in gradients. Indeed,
\be \varPi^{(0),[2]}_* =
      -C\ell^4(\widetilde{\grad}\ol{\omega})^\top
\ol{\bS}(\widetilde{\grad}\ol{\omega})
     -C'\ell^4 \ol{\bS} \bdots (\grad\ol{\omega}\bdot\grad)\widetilde{\ol{\bS}}
\lb{flux-NL2D-1}
\ee
with $C=[C_4^{(0)}-2(C_2^{(0)})^2]/32$ and $C'=C_4^{(0)}/16.$ Using
$\ben^\top\ol{\bS}\ben =-\ol{\bS}$ and (\ref{Diff-TildeS}), this can also be
written as
\be  \varPi^{(0),[2]}_* = C\ell^4 (\grad\ol{\omega})^\top
\ol{\bS}(\grad\ol{\omega})
       + 4C'\ell^4 \ol{\sigma}^2 (\grad\ol{\omega}\bdot\grad\ol{\alpha}).
\lb{flux-NL2D-2}
\ee
These are the only UV-local contributions to the energy flux at second-order.

It is important to determine the sign of these terms, on average, to see
whether they contribute to inverse cascade or direct cascade. In this respect,
note that the first term in (\ref{flux-NL2D-2}) is proportional to the negative
of the rate of vorticity-gradient stretching by the large-scale strain. That
is,
if one considers the equation for the large-scale vorticity gradient, then it
has
the form
\be \ol{D}_t |\grad\ol{\omega}|^2
    = -2(\grad\ol{\omega})^\top \ol{\bS}(\grad\ol{\omega}) + \cdots,
\lb{vort-grad-eq} \ee
where $\ol{D}_t=\partial_t+\ol{\bu}\bdot\grad$ and $\cdots$ denotes neglected
terms due to the turbulent stress. Thus, we see that the first term in
(\ref{flux-NL2D-2}) is negative (inverse cascade) precisely when
vorticity-gradients
are magnified, a connection already noted by \cite{Kraichnan76}. Equivalently,
inverse cascade requires the stretching direction ${\bmi e}_+^{(\sigma)}$ of
the strain
field to tend to be parallel to contour-lines of the large-scale vorticity.
Since we
have already seen that the small-scales induce a tensile stress along the
contour lines,
the stress and strain cooperate in this alignment and negative work is done by
the large scales against the small-scales. Equation (\ref{vort-grad-eq})
renders
the required alignment plausible, since components of the vorticity-gradient
parallel
to the squeezing direction will tend to grow, according to this equation. Note
that
this tendency might be moderated somewhat by the small-scale stress terms which
we have neglected in (\ref{vort-grad-eq}); cf. \cite{VanderBos02}.

The second term in (\ref{flux-NL2D-2}) will be negative precisely when
$\grad\ol{\omega}\bdot\grad\ol{\alpha}<0.$ This means that the strain frame
must counter-rotate against vorticity changes, i.e. rotate clockwise moving
in the direction of increasing vorticity. We do not have a direct dynamical
explanation for this tendency, analogous to the one we gave above
for vorticity-gradient stretching. On the other hand, we have found that there
is a simple kinematic relation between the rates of differential
strain-rotation
and vorticity-gradient stretching in 2D:
\be  \langle (\grad\ol{\omega})^\top \ol{\bS} (\grad\ol{\omega})\rangle
= -\langle \ol{\bS} \bdots (\grad\ol{\omega}\bdot\grad)\widetilde{\ol{\bS}}
\rangle
\lb{Betchov-2Da} \ee
or, equivalently,
\be  \langle (\grad\ol{\omega})^\top \ol{\bS} (\grad\ol{\omega})\rangle
     = 4\langle\ol{\sigma}^2 (\grad\ol{\omega}\bdot{\grad\ol{\alpha}})\rangle.
\lb{Betchov-2Db} \ee
Equation (\ref{Betchov-2Da}) [or (\ref{Betchov-2Db})] is an exact 2D analogue
of
the 3D relation of \cite{Betchov56}, and, like it, depends just on homogeneity
and
incompressibility of the velocity field. For a proof of the `2D Betchov
relation'
(\ref{Betchov-2Da}), see Appendix C. An important immediate consequence is that
differential strain counter-rotation and vorticity-gradient stretching must
occur
together, on average, while differential strain co-rotation is associated with
mean
shrinking of vorticity-gradients\footnote{Because it is purely kinematic, the
`2D Betchov relation' holds just as well in the enstrophy cascade range. As
discussed
in \cite{Ey01} and \cite{CEEWX03}, forward enstrophy-flux is also associated
with mean stretching of filtered vorticity-gradients. Thus, differential strain
counter-rotation must also occur, on average, in the enstrophy cascade.}.

The net energy flux from both terms in (\ref{flux-NL2D-2}) is always negative
(inverse cascade) when there is mean stretching of vorticity-gradients.
Because of the Betchov-like relation (\ref{Betchov-2Db}) it follows that
$\langle \varPi^{(0),[2]}_* \rangle = (C+C')\ell^4 \Gamma,$
where $\Gamma$ is the common average in (\ref{Betchov-2Db}) and
\be C+C' = \frac{1}{32}C_4^{(0)}+\frac{1}{16}[C_4^{(0)}-(C_2^{(0)})^2]\geq 0.
\lb{const-sum} \ee
To prove inequality (\ref{const-sum}), note that $C_4^{(0)}\geq 0$ by its
definition.
Furthermore,
\be C_2^{(0)}=\int_{|\br|\geq 1} d\br \,\,|\br|^2 G(\br)
    \leq \sqrt{\int_{|\br|\geq 1} d\br \,\,G(\br)
         \cdot \int_{|\br|\geq 1} d\br \,\,|\br|^4 G(\br)}
    \leq  \sqrt{C_4^{(0)}} \lb{C-S-ineq}
\ee
by the Cauchy-Schwartz inequality and normalization of $G$. This gives
(\ref{const-sum}). Thus, for {\it any} filter, the net flux is negative
when $\Gamma<0.$ The 2D Betchov relation furthermore gives the ratio of
contribution to inverse cascade of the two terms in (\ref{flux-NL2D-2}),
as $C/C'.$ For a Gaussian filter this ratio is $C/C'=(1/2)-(9/13)e^{-1/2}
\doteq 0.08,$ so that approximately 92.6\% of the mean of (\ref{flux-NL2D-2})
comes from differential strain-rotation and 7.4\% from vorticity-gradient
stretching.

\subsubsection{The Weakly UV-Local Terms}

The terms of the MSG expansion that are second-order in
gradients contribute to energy flux already from the strongly
UV-local modes. However, there are additional contributions at
second-order from all the other subscale modes. Here we shall
discuss the physical interpretation and significance of those.

In fact, the various terms that appear in the expressions for the 2D model
stress,
(\ref{MNL-stress-2D-diag}) and (\ref{MNL-stress-2D-off}), can be readily
understood.
The first term in (\ref{MNL-stress-2D-diag}), which is first-order in
gradients,
has already been discussed. In the next group of three 2nd-order terms, the
first
and last are both pressure contributions and do not contribute to energy flux.
However, the middle term is deviatoric and can give rise to flux. Using the
analogue of
(\ref{Diff-TildeS}),
\be (\grad\omega^{[k]}\bdot\grad)\widetilde{\bS}^{[k]}
   =-2(\grad\omega^{[k]}\bdot\grad\alpha^{[k]})\bS^{[k]}
    +(\grad\omega^{[k]}\bdot \grad\lambda^{[k]})\widetilde{\bS}^{[k]},
\lb{Diff-TildeS-k} \ee
this term can be split into two. The first is a Newtonian stress
$-2\nu_T^{[k]}\bS^{[k]}$
with an eddy-viscosity coefficient
\be \nu_T^{[k]}= (1/16)\ol{C}_4^{[k]}\ell^4_k(\grad\omega^{[k]}\bdot
                              \grad\alpha^{[k]})\ee
arising from differential strain-rotation at a length-scale
$\ell_k.$ The other term is of the `skew-Newtonian' form $\gamma_T^{[k]}
\widetilde{\bS}^{[k]}$ with skew-viscosity coefficient
\be \gamma_T^{[k]}= (1/16)\ol{C}_4^{[k]}\ell^4_k(\grad\omega^{[k]}
                            \bdot\grad\lambda^{[k]}) \ee
arising from {\it differential strain-magnification} at the same length-scale
$\ell_k.$
Note that we have defined the logarithm of the strain eigenvalue or magnitude
as
$\lambda^{[k]}=\ln\sigma^{[k]}.$ Since the velocity field in the inverse
cascade range
is monofractal with H\"{o}lder exponent $1/3$ (\cite{ParetTabeling98,Yakhot99,
Boffettaetal00}), it is not hard to see that both $\nu_T^{[k]}$ and
$\gamma_T^{[k]}$
are of order $O(\ell_k^{4/3}),$ as expected. \footnote{To show this, use the
formulas
$2\grad\alpha = \frac{{\sS}_{11}\grad {\sS}_{12}-{\sS}_{12}\grad {\sS}_{11}}
{{\sS}_{11}^2+{\sS}_{12}^2},
  \,\,\,\,\,\,\,\,
  \grad\lambda = \frac{{\sS}_{11}\grad {\sS}_{11}+{\sS}_{12}\grad {\sS}_{12}}
  {{\sS}_{11}^2+{\sS}_{12}^2}, $
and the general estimates from \cite{Eyink05} and (\cite{Eyink05a})}. The last
term in (\ref{MNL-stress-2D-diag}) represents a tensile stress of magnitude
$+\ol{C}_4^{[k]}\ell^4_k|\grad\omega^{[k]}|^2/32$ exerted along contour-lines
of the vorticity $\omega^{[k]}$ at length-scale $\ell_k$.

There remains the `fluctuation' contribution to the stress from
(\ref{MNL-stress-2D-off}). This can be best understood by summing over scales,
to give $\bu_*^{\prime\,(n,2)} = \widetilde{\grad}\psi^{(n)}_* $ with a
{\it fluctuation stream-function}
\be \psi^{(n)}_* = \frac{1}{4} \sum_{k=0}^n
\frac{\ol{C}_2^{[k]}}{\sqrt{N_k}}\,\,\ell^2_k\omega^{[k]}.
\lb{stream-fun} \ee
Note that the factor $1/\sqrt{N_k}$ reflects the cancellations that
are expected to occur in the space-integral for the contributions from modes
at length-scale $\ell_k$ (I). We see then, finally, that
$-\widetilde{\grad}\psi^{(n)}_*
\widetilde{\grad}\psi^{(n)}_*$ represents a contractile stress along the
streamlines
of $\psi^{(n)}_*$. This term opposes and, to some degree, cancels against the
tensile stress terms in (\ref{MNL-stress-2D-diag}) exerted along the
contour-lines
of $\omega^{[k]}$ for $k=1,...,n.$

If the model stress is substituted into formula (\ref{energy-flux}) for the
flux,
then there results:
\begin{eqnarray}
\varPi^{(n,2)}_* & = &
    \sum_{k=0}^n \left\{
   \frac{1}{2} \ol{C}_2^{[k]} \ell^2_k
\omega^{[k]}(\bS^{(0)}\bdots\,\widetilde{\bS}^{[k]})
    + \frac{1}{8} \ol{C}_4^{[k]}\ell^4_k
(\grad\omega^{[k]}\bdot\grad\alpha^{[k]})
        (\bS^{(0)}\bdots\,\bS^{[k]}) \right. \cr
  & & \,\,\,\,\,\,\,\,\,\,\,\,\,\left.
   -\frac{1}{16} \ol{C}_4^{[k]}\ell^4_k
(\grad\omega^{[k]}\bdot\grad\lambda^{[k]})
        (\bS^{(0)}\bdots\,\widetilde{\bS}^{[k]})
   + \frac{1}{32}\ol{C}^{[k]}_4 \ell^4_k \,
     (\grad\omega^{[k]})^\top\bS^{(0)}(\grad\omega^{[k]})
   \right\} \cr
  & &
\,\,\,\,\,\,\,\,\,\,\,\,\,\,\,\,\,\,\,\,\,\,
\,\,\,\,\,\,\,\,\,\,\,\,\,\,\,\,\,\,\,\,\,\,
\,\,\,\,\,\,\,\,\,\,\,\,-(\grad\psi^{(n)}_*)^\top\bS^{(0)}(\grad\psi^{(n)}_*)
\lb{MNL-flux-2D}
\end{eqnarray}
This is our final CSA expansion result for the energy flux in 2D. In addition
to the first-order term that appeared in (\ref{MNL-total-flux-1st-polar}),
there
are now second-order contributions arising from differential strain-rotation,
differential strain-magnification, and vorticity-gradient stretching. The final
term in (\ref{MNL-flux-2D}) is expected to be much smaller than the others,
because of the cancellations in space-averaging discussed above and additional
cancellations in the sum over scales in (\ref{stream-fun}). We expect
that the first four terms contribute to inverse cascade. For small $k,$
$\bS^{[k]}$ should be correlated to some degree with $\bS^{(0)},$ so that
the differential strain-rotation and vorticity-gradient stretching terms ought
to have negative mean-values, for similar reasons as the corresponding $k=0$
terms discussed earlier. Like the first-order `skew-Newtonian' term,
the differential strain-magnification term vanishes for $k=0$ and can
therefore be expected to be relatively smaller than the differential
strain-rotation term. It is interesting to note that the latter has
its sign determined by the quantity
$\grad\omega^{[k]}\bdot\grad\alpha^{[k]}\cos[2(\alpha^{[k]}-\alpha^{(0)})],$
closely related to the signed quantity
$\omega^{[k]}\sin[2(\alpha^{[k]}-\alpha^{(0)})]$
that appears in the first-order term. The final term in (\ref{MNL-flux-2D})
is the only one that we expect to have a positive mean (from vorticity-gradient
stretching), but we have already argued that that term will be considerably
smaller
in magnitude.

Note that the flux term in (\ref{MNL-flux-2D}) from scale $k$ gives at most
a fraction of order $2^{-2k/3}$ to the net energy flux. This agrees with
rigorous
locality estimates (\cite{Eyink05}). However, the actual contribution is likely
to be much smaller, since the correlations which produce the inverse energy
cascade
must weaken for $k\gg 1.$ If the small-scales are isotropic, then the mean
stress
$\btau^{[k]}$ from length-scale $\ell_k$ will satisfy:
\be \langle \tau_{ij}^{[k]}\rangle =
   \frac{1}{2} \langle {\rm Tr}\,[\btau^{[k]}]\rangle \delta_{ij},
    \,\,\,\,\,\,\,\,\,\,\mbox{for $k\gg 1$} .
\lb{iso-stress} \ee
In that case, if the large-scale strain $\bS^{(0)}$ and the stress contribution
$\btau^{[k]}$ are asymptotically independent for $k \gg 1,$ then their mean
contribution to the energy flux vanishes, since the deviatoric part of the
stress
is zero on average. The existence of an energy cascade requires a statistical
correlation between the large-scale strain and the small-scale stress
contributions from various scales, which becomes progressively weaker
for increasing $k.$

\section{Discussion}

The theoretical expression that we have developed here for the
turbulent stress yields many concrete testable predictions---both
qualitative and quantitative---for the 2D inverse energy cascade.
Foremost, we predict that strain-frames at small scales should
lag/lead those at large-scales, when the small-scale vorticity is
positive/negative. A spatial analogue of this effect is that the
strain eigenframes are predicted on average to rotate clockwise in
the direction of increasing vorticity (differential
counter-rotation). Likewise, we predict that there will be a
positive mean rate of stretching of vorticity-gradients. More
quantitatively, our final CSA-MSG formulas (\ref{MNL-stress-II}),
(\ref{MNL-stress-2D-diag}), (\ref{MNL-stress-2D-off}) for the stress
and (\ref{MNL-flux-2D}) for the flux may be compared in detail with
results obtained from experiment or simulation. If the model survives
such tests, then it may be a good point of departure for building
a practical LES modelling scheme of the 2D inverse energy cascade.

In our presentation above we have alluded only briefly to the dynamical
mechanisms that can produce the various correlations and alignments that
are postulated, e.g. based on the evolution equations of strain
orientation-angles
(\ref{alpha-n-dot}) and of vorticity-gradients (\ref{vort-grad-eq}). Many
of the mechanisms expected to operate in 2D have very close analogues in 3D.
Notice that vortex-stretching in 3D is a near relative of the vortex-thinning
mechanism in 2D, which we discussed in section 2.1.3. However, the
result is opposite, because the stretching process in 3D ``spins up''
the vortices and increases the kinetic energy in the small scales. Vorticity
contour-lines in 2D can also be expected to lengthen on the basis of the
same plausible statistical arguments that have been applied to vortex
lines or other material lines in 3D (\cite{Taylor38,Batchelor52,Cocke69}).
This already argues rather strongly for the stretching of vorticity-gradients
in 2D
incompressible turbulence and, via the Betchov-like relation
(\ref{Betchov-2Db}),
for differential rotation of strain counter to vorticity. On the other hand, in
3D
rather more detailed understanding is available through simple Lagrangian
models of
the evolution of velocity-gradients
(\cite{Vieillefosse82,Vieillefosse84,Cantwell92},
Chertkov, Pumir \& Shraiman (1999)). These phenomenological models have
provided plausible
dynamical explanations of the key alignments that are observed in DNS
(\cite{Ashurstetal87})
and experiment (\cite{Taoetal02}). Some of the difficulties in developing such
understanding of the inverse energy cascade can be appreciated by considering
the
exact equations in 2D for Lagrangian time-derivatives of the
velocity-gradients:
\be \left\{ \begin{array}{l}
            \ol{D}_t \ol{\omega}=0 \cr
            \ol{D}_t\ol{\sS}_{ij}=\frac{1}{2}(\bigtriangleup \ol{p})\delta_{ij}
                             - \partial^2_{ij}\ol{p}
           \end{array} \right. \lb{2D-ugrad-eq} \ee
Here we have considered separately the evolution of the vorticity and strain.
We have
also neglected the contribution of turbulent stresses to the evolution of
filtered
gradients, which may  be an important feedback interaction with small-scales
(\cite{VanderBos02}). The equations (\ref{2D-ugrad-eq}) lack the local
self-stretching
terms which play the key role in the analogous 3D equations. In fact, the
Lagrangian
evolution in (\ref{2D-ugrad-eq}) is entirely trivial except for the pressure
hessian in
the equation for the strain and the latter must play an essential role in the
production
of strain orientation alignments. More sophistication in the modeling of
pressure is
therefore likely to be required than in the 3D case
(\cite{Vieillefosse82,Vieillefosse84,
Cantwell92,Chertkovetal99}). Furthermore, we have seen that in the 2D inverse
cascade,
both higher-order gradient and multi-scale effects are important. Thus, it
remains a
challenge to develop a detailed dynamical understanding of the 2D inverse
energy cascade.

\vspace{.4in}

\noindent {\bf Acknowledgements:} I wish to thank S. Chen, B. Ecke,
M. K. Rivera, M.-P. Wang, and Z. Xiao for a very fruitful collaboration
on 2D turbulence which helped to stimulate the development of the
present theory. I would also like to thank C. Meneveau and E. Vishniac
for helpful discussions. This work was supported in part by NSF grant
\# ASE-0428325.

\newpage

\appendix

\section{Dynamical Equation for the Strain Orientation}

It is easy to see from the `polar' representation (\ref{stress-polar})
of the strain $\ol{\bS} $ that $2\ol{\alpha}=\arctan(\ol{\sS}_{12}
/\ol{\sS}_{11}).$ Since also $\ol{\sigma}^2=\ol{\sS}_{12}^2+\ol{\sS}_{11}^2,$
the Lagrangian derivative may be written as
\be 2\ol{\sigma}^2 \ol{D}_t\ol{\alpha} =
\ol{\sS}_{11}(\ol{D}_t\ol{\sS}_{12})-\ol{\sS}_{12}(\ol{D}_t\ol{\sS}_{11})
\lb{alpha-eq} \ee
We can evaluate the time rate of change from the equation (\ref{2D-ugrad-eq})
for the filtered strain, which neglects the contribution from turbulent stress.
Substituting into (\ref{alpha-eq}) we get
\begin{eqnarray}
2\ol{\sigma}^2\ol{D}_t\ol{\alpha} & = &
\frac{\partial\ol{u}}{\partial x}\left[-\frac{\partial^2\ol{p}}{\partial
x\partial y}\right]
-\frac{1}{2}\left(\frac{\partial\ol{u}}{\partial
y}+\frac{\partial\ol{v}}{\partial x}\right)
\left[\frac{1}{2}\bigtriangleup\ol{p}-\frac{\partial^2\ol{p}}{\partial
x^2}\right] \cr
& = & \frac{1}{2}\left(\frac{\partial\ol{u}}{\partial
y}+\frac{\partial\ol{v}}{\partial x}\right)
      \frac{\partial^2\ol{p}}{\partial x^2}
      -\frac{1}{2}\left(\frac{\partial\ol{u}}{\partial
x}-\frac{\partial\ol{v}}{\partial y}\right)
      \frac{\partial^2\ol{p}}{\partial x\partial y}
      -\frac{1}{4}\left(\frac{\partial\ol{u}}{\partial
y}+\frac{\partial\ol{v}}{\partial x}\right)
      \bigtriangleup\ol{p}
\lb{alpha-eq-first}
\end{eqnarray}
\noindent where we used incompressibility in the last line and also to derive
the next identity:
\be \frac{\partial\ol{u}}{\partial y}\frac{\partial^2\ol{p}}{\partial x^2}
    -\frac{\partial\ol{v}}{\partial x}\frac{\partial^2\ol{p}}{\partial y^2}
    -\left(\frac{\partial\ol{u}}{\partial x}-\frac{\partial\ol{v}}{\partial
y}\right)
     \frac{\partial^2\ol{p}}{\partial x\partial y}
    = \frac{\partial}{\partial x}
    \left(\frac{\partial\ol{p}}{\partial x}\frac{\partial\ol{u}}{\partial y}
         -\frac{\partial\ol{p}}{\partial y}\frac{\partial\ol{u}}{\partial
x}\right)
      + \frac{\partial}{\partial y}
    \left(\frac{\partial\ol{p}}{\partial x}\frac{\partial\ol{v}}{\partial y}
         -\frac{\partial\ol{p}}{\partial y}\frac{\partial\ol{v}}{\partial
x}\right)
\lb{main-alpha-id}
\ee
If (\ref{main-alpha-id}) is used in (\ref{alpha-eq-first}) to eliminate the
mixed partial
derivative of pressure, then one obtains
\be
2\ol{\sigma}^2\ol{D}_t\ol{\alpha} =
\frac{1}{4}\left(\frac{\partial\ol{v}}{\partial
x}-\frac{\partial\ol{u}}{\partial y}\right)
      \bigtriangleup\ol{p}
      + \frac{1}{2} \frac{\partial}{\partial x}
    \left(\frac{\partial\ol{p}}{\partial x}\frac{\partial\ol{u}}{\partial y}
         -\frac{\partial\ol{p}}{\partial y}\frac{\partial\ol{u}}{\partial
x}\right)
      + \frac{1}{2} \frac{\partial}{\partial y}
    \left(\frac{\partial\ol{p}}{\partial x}\frac{\partial\ol{v}}{\partial y}
         -\frac{\partial\ol{p}}{\partial y}\frac{\partial\ol{v}}{\partial
x}\right)
\lb{alpha-eq-last}\ee
This last equation is equivalent to
(\ref{alpha-n-dot}),(\ref{press-hess-n}),(\ref{K-curr-n})
in the text.

\section{Vortex-Thinning and Negative Eddy-Viscosity}

Some while ago, Kraichnan proposed a physical mechanism to explain the origin
of negative
eddy-viscosities in 2D (see \cite{Kraichnan76}, Section 5.) For this purpose he
employed
a simplified model of small-scale vortex wavepackets in a uniform, large-scale
straining
field. His aim was to understand the asymptotic effect of the small-scales on
much larger
scales, and not to give an account of the inverse energy cascade by scale-local
interactions.
Nevertheless, his ideas turn out to have much in common with our theory of the
local
cascade interactions. The model proposed by us in section 2.1.3 to explain the
stress
proportional to skew-strain is just a slight modification of Kraichnan's.
Furthermore,
his mechanism of `negative viscosity' is essentially identical with that we
found
in the last term of our model stress, equation (\ref{NL-model-2D}), which
corresponds
to a tensile stress along vorticity-contour lines. Here we shall review the
calculation
of \cite{Kraichnan76}, in order to make more clear its relation to the present
theory.

Kraichnan's model of the small-scales was a Gaussian wave-packet of vorticity
---called a `blob'---or an `assembly of uncorrelated blobs'
(\cite{Kraichnan76}).
The stream function of each blob was taken to have the form
\be \left.
    \begin{array}{l}
    \psi(\bx) = k^{-2} f(\bx) \cos(kx_2) \cr
    f(\bx) = \exp(-\frac{1}{2}(x_1^2+x_2^2)/D^2)
    \end{array} \right\} \ee
where $f$ is a Gaussian envelope function with a standard deviation $\sim D$
that is modulated by an oscillating cosine with wavevector $\bk$ pointing in
the
vertical ${\bmi e}_2$-direction. A basic assumption is that $kD\gg 1,$ so that
the wavenumber of the packet can be regarded as nearly sharp. Calculating the
small-scale velocity field from $\bu=-\widetilde{\grad}\psi,$  it is not hard
to show that the leading component of the velocity is
\be u_1 \sim k^{-1} f(\bx) \sin(kx_2) \lb{Kr76-vel} \ee
and of the vorticity-gradient is
\be (\grad\omega)_2 \sim  -k f(\bx) \sin(kx_2) \lb{Kr76-vort} \ee
asymptotically for $kD\gg 1.$ Cf. Eq.(5.4) in \cite{Kraichnan76}. Thus, the
dominant
component of the total stress
$\bT =  \int \btau= \int \bu\bu$ is
\be \sT_{11} = k^{-2} \int dx_1\int dx_2 \,\exp(-(x_1^2+x_2^2)/D^2)
                               \sin^2(kx_2)
                             \sim \pi D^2/2k^2 \lb{Kr76-stress} \ee
for $kD\gg 1.$ That is, the dominant stress is positive, or tensile, and
exerted
along the horizontal direction ${\bmi e}_1$. This is perpendicular to the
direction of
the vorticity-gradient ${\bmi e}_2,$ or along the direction of the
vorticity-contours.
Thus, Kraichnan's `blob model' leads to a result in agreement with our general
conclusion.

As a model of the large-scales, Kraichnan took a uniform strain field
\be  \bS         =   a \left(\begin{array}{cc}
                           \cos(2\phi) & \sin(2\phi) \cr
                           \sin(2\phi) & -\cos(2\phi)
                    \end{array}\right) \lb{Kr76-strain} \ee
with eigenvalues $\pm a$ and eigenframe oriented at an angle $\phi$ with
respect
to the fixed coordinate frame. The stream function corresponding to this
large-scale
field is just $V(\bx)=\frac{1}{2}\bx^\top\widetilde{\bS}\bx.$ Actually,
Kraichnan
kept the strain fixed with frame axes along the coordinate directions and
instead
rotated the wavenumber of the small-scale blob, as $\bk=k [ {\bmi e}_1 \sin\phi
+ {\bmi e}_2 \cos\phi ];$ Eq.(5.14) in \cite{Kraichnan76}. This is physically
more
natural, if one thinks of the small-scales as isotropic and the large-scales
as having fixed anisotropy. However, it is mathematically equivalent to rotate
the
strain and it relates more easily to our analysis in the text.

\cite{Kraichnan76} worked out in detail the energy balance for
his simple two-scale model of the velocity field. The initial energy in the
small-scales is
\be E=\frac{1}{2}\int |\bu|^2 \sim (1/2)\sT_{11}\sim \pi D^2/4k^2
                   \lb{Kr76-energy} \ee
The effect of the straining field on the small-scale wavevector is to change
its
magnitude by
\be dk^2/dt = -2 \bk^\top \bS \bk= 2 a \cos(2\phi) k^2. \lb{Kr76-dkdt} \ee
Thus, Kraichnan concluded that, to leading order,
\be  (dE/dt)_{t=0}=-\pi a D^2\cos(2\phi)/2k^2. \lb{Kr76-dEdt} \ee
Cf. Eq.(5.8) in \cite{Kraichnan76} for the case that $k=1$ and $\phi=0.$ This
reduction in energy of the small-scale blob is a consequence of the transfer of
its enstrophy to higher wavenumber.

Kraichnan showed further that the energy budget was maintained by a deposit
into the `interaction energy' $\int \bv\bdot\bu$ between the large-scale
and small-scale velocity fields. In his calculation he rewrote the interaction
energy as $\int V \omega,$ in terms of the large-scale streamfunction $V$ and
small-scale-vorticity $\omega$, and considered the nonlinear self-interaction
of
the latter. He found that the small-scale vorticity field set up a secondary
flow of four equal-strength vortices with alternating signs of circulation
which, for $\phi=0,$ reinforced the large-scale strain. In his own words:
\vspace{.1in}
\begin{quote}
`If a small-scale motion has the form of a compact blob of vorticity,
or an assembly of uncorrelated blobs, a steady straining will eventually
draw a typical blob out into an elongated shape, with corresponding
thinning and increase of typical wavenumber. The typical result will
be a decrease of the kinetic energy of the small-scale motion and a
corresponding reinforcement of the straining field....'
\end{quote}
\vspace{.1in}
In this way, the energy loss from the small-scales that is observed in
(\ref{Kr76-dEdt}) can traced to a transfer of equal size into the interaction
energy between large-scales and small-scales.

This transfer can be shown to be equivalent to the scale-to-scale energy flux
that we defined in (\ref{energy-flux}). Indeed, using the fact that the
large-scale velocity $\bv$ is stationary and its velocity-gradient $\grad\bv$
is uniform, we find that
\be (dE/dt)_{t=0} = -\frac{d}{dt}\int \bv\bdot\bu =
     \int \bv\bdot[\grad\bdot(\bu\bu)] = -\int (\grad\bv)\bdots \bu\bu
     = -\bS \bdots \bT.  \lb{Kr76-flux} \ee
This is the area-integral of the quantity that appears in (\ref{energy-flux}).
We can use this expression to easily verify the energy balance result from
\cite{Kraichnan76}. Substituting the stress from (\ref{Kr76-stress})
and the strain from (\ref{Kr76-strain}), one gets $(dE/dt)_{t=0}=
-\pi a D^2\cos(2\phi)/2k^2,$ in agreement with (\ref{Kr76-dEdt}).
Note that the flux is negative and the small-scales lose energy only if
$|\phi|<\pi/4,$ whereas the flux is positive for $\pi/4<|\phi|<\pi/2$.
If one assumes that the angle $\phi$ is random with an isotropic
distribution and $\bk=k{\bmi e}_2$ is fixed, then the average flux is
$\langle(dE/dt)_{t=0}\rangle_{{\rm ang}}=0.$ \cite{Kraichnan76}
had already noted this result and established its consistency with
the mean growth of small-scale wavenumber magnitude or, equivalently,
the mean stretching of small-scale vorticity-gradients. As we discussed
around our equation (\ref{iso-stress}), a mean energy flux under
isotropic conditions requires statistical correlations between disparate
scales. In Kraichnan's case where he assumed a very wide separation between
the two scales of motion, it was realistic to assume negligible correlations
and thus zero net transfer. However, this is an unrealistic assumption
in the context of a local energy cascade, where the stress and strain
in (\ref{energy-flux}) get most of their contributions from adjacent
scales (\cite{Eyink05}) and are highly correlated.

It is interesting that the mechanism that Kraichnan identified as acting
between distant scales can also be identified with several of the mechanisms
that we have found in our analysis of local cascade interactions.
Note that in Kraichnan's vortex-blob model
\be (\grad\omega)^\top\bS(\grad\omega) = -a k^2
f^2(\bx)\sin^2(kx_2)\cos(2\phi),
\lb{penult-Kr76} \ee
using (\ref{Kr76-vort}) and (\ref{Kr76-strain}). Integrated over space, this
yields
\be \int (\grad\omega)^\top\bS(\grad\omega) = -\pi a (Dk)^2 \cos(2\phi)/2,
\lb{ultimate-Kr76} \ee
to leading order for $Dk\gg 1.$ Thus, we get agreement of (\ref{Kr76-dEdt})
with the fourth term in our formula $\varPi^{(n,2)}_*$ for the energy flux,
equation (\ref{MNL-flux-2D}), by taking $\ell_k=1/k$ there. The second term in
(\ref{MNL-flux-2D}) corresponding to differential strain rotation is zero
in the vortex-blob model because the orientations of the strain-fields
(both large-scale and small-scale) are uniform in space. However, we can
equally
well understand the energy flux in the blob model based upon the first term in
(\ref{MNL-flux-2D}) [the same as (\ref{MNL-flux-1st})] that corresponds to
relative rotation of strain at disparate scales. Indeed, in the blob-model,
the vorticity is
\be \omega(\bx) \sim -f(\bx) \cos(kx_2) \lb{blob-vorticity} \ee
and the small-scale strain of the blob is
\be  \bS'(\bx)=\frac{1}{2}\omega(\bx) \left(\begin{array}{cc}
                            0 & -1 \cr
                            -1 & 0
                    \end{array}\right) \lb{blob-strain} \ee
to leading order. Thus it is not hard to calculate that
\be \omega \bS\bdots\widetilde{\bS}' = -a f^2(\bx)\cos^2(kx_2)\cos(2\phi)
                                \lb{blob-first} \ee
and integrated over space this gives also
\be \int \omega \bS\bdots\widetilde{\bS}' = -\pi a D^2 \cos(2\phi)/2
\lb{blob-first-avrg} \ee
to leading order for $Dk\gg 1.$ Multiplying (\ref{blob-first-avrg}) by
$\ell_k^2=1/k^2,$ we get also agreement of (\ref{Kr76-dEdt}) with formula
(\ref{MNL-flux-1st}). It is intriguing to note that, before averaging over
space,
the two contributions from (\ref{penult-Kr76}) and (\ref{blob-first}) are
exactly out of phase. It is another simple exercise to verify that the
third term in (\ref{MNL-flux-2D}), from differential strain-magnification,
is also non-zero in the blob model and gives a contribution of the same sort.

Thus, it is clear that most of the terms in our CSA-MSG formula
(\ref{MNL-flux-2D}) are represented in Kraichnan's blob model,
in particular, the flux from skew-strain, from differential
strain-magnification, and from vorticity-gradient stretching. All
of these can be produced by a single mechanism of
`vortex-thinning'. Our somewhat simpler model of vortex patches in
section 2.1.3 also illustrates these same flux terms, except in
the case of constant-vorticity patches, for which only the flux
from skew-strain survives. The increase in wavenumber that was
considered by Kraichnan in his blob model and the asymptotics
$Dk\gg 1$ play no essential role in the skew-strain mechanism.
Indeed, note that (\ref{blob-vorticity})-(\ref{blob-first-avrg})
for the blob model all have non-vanishing values at $k=0,$ whereas
(\ref{penult-Kr76})-(\ref{ultimate-Kr76}) tend to zero as
$k\rightarrow 0.$


\section{2D Betchov Relation}

For any incompressible or solenoidal field $\bu$ in 2D we can define a
corresponding
`strain' $\sS_{ij}^{(u)}$ and `vorticity' $\omega^{(u)}$ via
\be \frac{\partial u_i}{\partial x_j} = u_{i,j}= \sS_{ij}^{(u)}-\frac{1}{2}
                           \epsilon_{ij}\omega^{(u)}. \lb{gradu-S-om-app} \ee
Observe our notation for partial derivative with respect to $x_j,$ indicated
by subscript $j$ preceded by a comma. Likewise, we write $\partial^2
u_i/\partial x_j
\partial x_k=u_{i,jk},$ etc. Using these notations and definitions, the first
step
in the derivation of the 2D Betchov relation is the following identity:
\begin{eqnarray}
\partial_l[\sS_{ij}^{(u)}v_{i,k}w_{j,kl}]
-\partial_k[\sS_{ij}^{(u)}v_{i,k}w_{j,ll}]
   & = & \underbrace{\sS_{ij}^{(u)}v_{i,kl}w_{j,kl}}
         - \underbrace{\sS_{ij}^{(u)}v_{i,kk}w_{j,ll}} \cr
   &   &  \,\,\,\,\,\,\,\,\,\,\,\,\,\,\,\,\,
          \mbox{\circle{8}}\!\!\!\!\!\!_1
\,\,\,\,\,\,\,\,\,\,\,\,\,\,\,\,\,\,\,\,\,\,\,\,\,\,\,\,\,\,\,\,\,\,\,\,\,\,\,
          \mbox{\circle{8}}\!\!\!\!\!\!_2 \cr
   &   & {\,} \cr
  &   & + \underbrace{\sS_{ij,l}^{(u)}v_{i,k}w_{j,kl}}
         -\underbrace{\sS_{ij,k}^{(u)}v_{i,k}w_{j,ll}} \cr
  &   & \,\,\,\,\,\,\,\,\,\,\,\,\,\,\,\,\,\,\,\,\,
          \mbox{\circle{8}}\!\!\!\!\!\!_3
\,\,\,\,\,\,\,\,\,\,\,\,\,\,\,\,\,\,\,\,\,\,\,\,\,\,\,\,\,\,\,\,\,\,\,\,\,\,\,
          \mbox{\circle{8}}\!\!\!\!\!\!_4
\lb{2D-Betchov-1st-id} \end{eqnarray}
Here $\bu,\bv,\bw$ are all incompressible fields. The identity
(\ref{2D-Betchov-1st-id})
follows straightforwardly from the product of rule of differentiation.

The terms labelled ${\!\,}^{\,\,\,\mbox{\circle{8}}\!\!\!\!\!\!\!_1\,\,},$
${\!\,}^{\,\,\,\mbox{\circle{8}}\!\!\!\!\!\!\!_2\,\,},$ and
${\!\,}^{\,\,\,\mbox{\circle{8}}\!\!\!\!\!\!\!_3\,\,}$ are easily calculated
by the substitutions $v_{i,j}= \sS_{ij}^{(v)}-(1/2)\epsilon_{ij}\omega^{(v)},
v_{i,jk}= \sS_{ij,k}^{(v)}-(1/2)\epsilon_{ij}\partial_k\omega^{(v)},$ and
$v_{i,kk}=-\epsilon_{il}\partial_l\omega^{(v)},$ and similar substitutions
for the field $\bw.$ The term
${\!\,}^{\,\,\,\mbox{\circle{8}}\!\!\!\!\!\!\!_1\,\,}$ becomes
\be \sS_{ij}^{(u)}v_{i,kl}w_{j,kl} = \frac{1}{2}S_{ij}^{(u)}\left(
    \widetilde{\sS}_{ij,l}^{(v)}\partial_l\omega^{(w)} +
    \widetilde{\sS}_{ij,l}^{(w)}\partial_l\omega^{(v)}\right),
\lb{B-1} \ee
${\!\,}^{\,\,\,\mbox{\circle{8}}\!\!\!\!\!\!\!_2\,\,}$ becomes
\be -\sS_{ij}^{(u)}v_{i,kk}w_{j,ll} =  \sS_{ij}^{(u)} \partial_i\omega^{(v)}
     \partial_j\omega^{(w)}, \lb{B-2} \ee
and ${\!\,}^{\,\,\,\mbox{\circle{8}}\!\!\!\!\!\!\!_3\,\,}$ becomes
\be \sS_{ij,l}^{(u)}v_{i,k}w_{j,kl} =
-\frac{1}{2}\sS_{ik}^{(v)}\widetilde{\sS}_{ik,l}^{(u)}\partial_l\omega^{(w)}
    +\frac{1}{2}\sS_{ij,l}^{(u)}\widetilde{\sS}_{ij,l}^{(w)}\omega^{(v)}.
\lb{B-3} \ee
The last term ${\!\,}^{\,\,\,\mbox{\circle{8}}\!\!\!\!\!\!\!_4\,\,}$ requires
as an additional step to use the identity
\be \sS_{ij,k}^{(u)}-\frac{1}{2}\epsilon_{ij}\partial_k\omega^{(u)}
    = u_{i,jk} = u_{i,kj}
    = \sS_{ik,j}^{(u)}-\frac{1}{2}\epsilon_{ik}\partial_j\omega^{(u)}
\ee
to replace $\sS_{ij,k}^{(u)}$ by  $\sS_{ik,j}^{(u)}$. Then using the same
substitutions as for the other three terms,
 ${\!\,}^{\,\,\,\mbox{\circle{8}}\!\!\!\!\!\!\!_4\,\,}$ becomes
\be -\sS_{ij,k}^{(u)}v_{i,k}w_{j,ll} =
    \widetilde{\sS}_{ik,j}^{(u)}\sS_{ik}^{(v)}\partial_j\omega^{(w)}
    +\frac{1}{2}\sS_{jk}^{(v)}\partial_k\omega^{(u)}\partial_j\omega^{(w)}
    + \frac{1}{4}\epsilon_{kj}\partial_k\omega^{(u)}\partial_j\omega^{(w)}
      \omega^{(v)}
\ee

We are now able to sum the contributions from all four terms,
${\!\,}^{\,\,\,\mbox{\circle{8}}\!\!\!\!\!\!\!_1\,\,},$
${\!\,}^{\,\,\,\mbox{\circle{8}}\!\!\!\!\!\!\!_2\,\,},$
${\!\,}^{\,\,\,\mbox{\circle{8}}\!\!\!\!\!\!\!_3\,\,},$ and
${\!\,}^{\,\,\,\mbox{\circle{8}}\!\!\!\!\!\!\!_4\,\,}.$ In order to simplify
the
result, it is helpful to define the quantity
\be \sT_{ij}^{(u,v)}= \partial_i\omega^{(u)}\partial_j\omega^{(v)}
          + \widetilde{\sS}_{ij,k}^{(u)}\partial_k\omega^{(v)}. \ee
Then the sum of the four terms yields, after some elementary algebra,
\begin{eqnarray}
\sS_{ij}^{(v)}\sT_{ij}^{(u,w)}+\sS_{ij}^{(u)}
\sT_{ij}^{(v,w)}+\sS_{ij}^{(u)}\sT_{ij}^{(w,v)}
   & = & \epsilon_{ij} [\sS_{ik,l}^{(u)}\sS_{jk,l}^{(w)}-
              \partial_i\omega^{(u)}\partial_j\omega^{(w)}]\omega^{(v)} \cr
  \,&  & -2\partial_k[\sS_{ij}^{(u)}v_{i,k}w_{j,ll}] +
2\partial_l[\sS_{ij}^{(u)}v_{i,k}w_{j,kl}]
\lb{2D-Betchov-uvw-part}
\end{eqnarray}
Observe that the first term on the righthand of (\ref{2D-Betchov-uvw-part}) is
antisymmetric under the interchange $\bu\leftrightarrow\bw.$ Thus, if we
symmetrize
(\ref{2D-Betchov-uvw-part}) in $\bu$ and $\bw,$ we get
\be \sS_{ij}^{(u)}\sT_{ij}^{(v,w)}+ {\rm perm.}
                       = {\rm div}\,[\cdots] \ee
where the sum on the lefthand side is over all six permutations of
$\bu,\bv,\bw$
and ${\rm div}\,[\cdots]$ on the righthand side indicates a total divergence.
It therefore follows that
\be \langle \sS_{ij}^{(u)}\sT_{ij}^{(v,w)}\rangle + {\rm perm.} = 0
\lb{2D-Betchov-uvw}
\ee
where $\langle\cdots\rangle$ denotes either an average over a homogeneous
ensemble
or a space-average with boundary conditions that permit integrations by parts
(e.g. periodic). We call the relation (\ref{2D-Betchov-uvw}) the {\it
generalized
Betchov identity in 2D}. Setting $\bu=\bv=\bw$ gives
\be \langle \sS_{ij}^{(u)}\sT_{ij}^{(u,u)}\rangle = 0  \lb{2D-Betchov-app-I}
\ee
where $\sS_{ij}^{(u)}$ and $\sT_{ij}^{(u,u)}$ are now constructed from the
field $\bu$
alone. Equation (\ref{2D-Betchov-app-I}) is equivalent to the 2D Betchov
relation
(\ref{Betchov-2Da}) or (\ref{Betchov-2Db}) stated in the text.

\end{document}